\documentclass[journal]{IEEEtran}
\usepackage{tikz}
\usepackage[mode=buildnew]{standalone}
\usepackage{pgfplots,siunitx}
\pgfplotsset{compat=newest}
\pgfplotsset{width=7cm,compat=1.3}
\usepackage{graphicx}
\usepackage{algpseudocode}

\pgfplotsset{compat=newest}%
\usetikzlibrary{positioning,matrix,shapes.multipart,shapes.misc,spy}%

\usetikzlibrary{spy}

\definecolor{lines-1}{RGB}{228,26,28}
\definecolor{lines-2}{RGB}{55,126,184}
\definecolor{lines-3}{RGB}{77,175,74}
\definecolor{lines-4}{RGB}{152,78,163}
\definecolor{lines-5}{RGB}{255,127,0}
\definecolor{lines-6}{RGB}{153,153,153}
\definecolor{lines-7}{RGB}{166,86,40}
\definecolor{lines-8}{RGB}{247,129,191}
\definecolor{lines-9}{RGB}{255,255,51}
\pgfplotscreateplotcyclelist{myCycleList}{
	color=lines-1,semithick,mark=o,mark size=2,mark repeat=1\\%
	color=lines-2,semithick,mark=star,mark size=3,mark repeat=1\\%
	color=lines-3,semithick,mark=square,mark size=2,mark repeat=1\\%
	color=lines-4,semithick,mark=diamond,mark size=3,mark repeat=1\\%
	color=lines-5,semithick,mark=triangle,mark size=3,mark repeat=1\\%
	color=lines-6,semithick,mark=Mercedes star,mark size=3,mark repeat=1\\%
	color=lines-7,semithick,mark=|,mark size=3,mark repeat=1\\%
	color=lines-8,semithick,mark=x,mark size=3,mark repeat=1\\%
}
\pgfplotsset{
	compat=1.14,
	width =\columnwidth, 
	height=.8\columnwidth,
	ylabel absolute, ylabel style={yshift=-0.2cm},
	xlabel absolute, xlabel style={yshift=0.2cm},
	label style={font=\normalsize},
	tick label style={font=\scriptsize},
	legend style={font=\footnotesize,cells={align=left}},
	grid=both,
	minor grid style={dotted},
}

\usepackage[font=small]{caption}

\usepackage{multirow}
\usepackage{graphicx}
\usepackage{cite}
\usepackage{balance}
\usepackage{color}
\usepackage{subcaption}
\usepackage{diagbox} 
\usepackage{algorithm} 
\usepackage{amssymb}
\usepackage{amsfonts}
\usepackage[utf8]{inputenc} 
\usepackage[T1]{fontenc}
\usepackage{url}
\usepackage{ifthen}
\usepackage{cite}
\usepackage{gensymb}
\usepackage[cmex10]{amsmath} 

\newcommand{\R}{\mathbb{R}}
\newcommand{\Z}{\mathbb{Z}}

\newcommand{\B}{\mathcal{B}}

\newcommand{\U}{\mathcal{U}}
\newcommand{\Q}{\mathcal{Q}}

\newcommand{\D}{\mathcal{D}}

\newcommand{\ba}{\boldsymbol{a}}
\newcommand{\bb}{\boldsymbol{b}}
\newcommand{\bc}{\boldsymbol{c}}

\newcommand{\be}{\boldsymbol{e}}
\newcommand{\bh}{\boldsymbol{h}}
\newcommand{\bl}{\boldsymbol{l}}

\newcommand{\bs}{\boldsymbol{s}}
\newcommand{\bt}{\boldsymbol{t}}
\newcommand{\bu}{\boldsymbol{u}}
\newcommand{\bv}{\boldsymbol{v}}
\newcommand{\bx}{\boldsymbol{x}}
\newcommand{\by}{\boldsymbol{y}}
\newcommand{\bz}{\boldsymbol{z}}

\newcommand{\bB}{\boldsymbol{B}}
\newcommand{\bC}{\boldsymbol{C}}
\newcommand{\bG}{\boldsymbol{G}}
\newcommand{\bGs}{\boldsymbol{G}_\mathrm{s}}

\newcommand{\bJ}{\boldsymbol{J}}

\newcommand{\bR}{\boldsymbol{R}}
\newcommand{\bX}{\boldsymbol{X}}
\newcommand{\bY}{\boldsymbol{Y}}
\newcommand{\bzero}{\boldsymbol{0}}
\newcommand{\bone}{\boldsymbol{1}}
\newcommand{\blambda}{\boldsymbol{\lambda}}

\newcommand{\Lambdas}{\Lambda_\mathrm{s}}
\newcommand{\Lambdac}{\Lambda_\mathrm{c}}
\newcommand{\Rc}{R_\mathrm{c}}

\DeclareMathOperator*{\argmin}{arg\,min} 

\ifCLASSINFOpdf
\else
\fi
\begin{document}

%
\title{Coded Modulation Schemes\\ for Voronoi Constellations}

%
%

\author{Shen~Li,
        Ali~Mirani,
        Magnus~Karlsson,~\IEEEmembership{Fellow,~IEEE,}~\IEEEmembership{Fellow,~OSA,}
        and~Erik~Agrell,~\IEEEmembership{Fellow,~IEEE}
\thanks{This research was funded in part by the Swedish Research Council (VR) under grants no. 2017-03702 and no. 2021-03709 and the Knut and Alice Wallenberg Foundation under grant no. 2018.0090.}
\thanks{S. Li and E. Agrell are with the Department
of Electrical Engineering, Chalmers University of Technology, 412 96 Gothenburg, Sweden. e-mail: shenl@chalmers.se.}
\thanks{A. Mirani was with the Department
of Microtechnology and Nanoscience, Chalmers University of Technology, 412 96 Gothenburg, Sweden and is now with Ericsson AB, 417 56, Gothenburg, Sweden.}
\thanks{M. Karlsson is with the Department
of Microtechnology and Nanoscience, Chalmers University of Technology, 412 96 Gothenburg.}
\thanks{Manuscript received July xx, 2023; revised xx xx, 2023.}
}

%
%

\markboth{Draft, July 27, 2023}%
{Shell \MakeLowercase{\textit{et al.}}: Bare Demo of IEEEtran.cls for IEEE Journals}
%



\maketitle

\begin{abstract}
Multidimensional Voronoi constellations (VCs) are shown to be more power-efficient than quadrature amplitude modulation (QAM) formats given the same uncoded bit error rate, and also have higher achievable information rates. However, a coded modulation scheme to sustain these gains after forward error correction (FEC) coding is still lacking. This paper designs coded modulation schemes with soft-decision FEC codes for VCs, including bit-interleaved coded modulation (BICM) and multilevel coded modulation (MLCM), together with three bit-to-integer mapping algorithms and log-likelihood ratio calculation algorithms. Simulation results show that VCs can achieve up to 1.84 dB signal-to-noise ratio (SNR) gains over QAM with BICM, and up to 0.99 dB SNR gains over QAM with MLCM for the additive white Gaussian noise channel, with a surprisingly low complexity.
\end{abstract}


%
\IEEEpeerreviewmaketitle

\section{Introduction}\label{sec:Intro}

\IEEEPARstart{A}{dvanced} multidimensional (MD) modulation formats are designed to have larger minimum Euclidean distance at the same average symbol energy than traditional two-dimensional (2D) quadrature amplitude modulation (QAM) formats. MD Voronoi constellations (VCs) are such a structured modulation format, comprising a coding lattice and a shaping lattice, the latter being a sublattice of the coding lattice \cite{conway83,forney89b}. The coding lattice determines 
how constellation points are packed, resulting in a coding gain over the cubic packing. The shaping lattice of VCs determines the boundary shape of the constellation, achieving a shaping gain over a hypercubic boundary. When applying soft-decision (SD) forward error correction (FEC) codes to VCs, the coding gain of FEC coding might fully or partially cover the coding gain of VCs. On the other hand, the shaping gain, which comes from improved signal distribution and is asymptotically 1.53 dB over QAM for the average power-constrained additive white Gaussian noise (AWGN) channel, cannot be realized by FEC coding.

VCs can have low-complexity encoding and decoding algorithms, i.e., mapping integers to constellation points and vice versa \cite{conway83,conway82decoding,feng13,ferdinandTWC,kurkoski18}, which entirely avoid the need to store and process all constellation points individually in the transmitter and receiver. VCs have shown better bit error rate (BER) performance than Gray-labeled QAM in uncoded systems \cite{ourISIT,ourjlt,alijlt20,aliecoc21,aliecoc22}. Mutual information (MI) and generalized mutual information (GMI) have also been studied for VCs in \cite{ourTC,ourjlt}, showing high gains over QAM. 

In modern communication systems, SD FEC codes are usually used to provide significant power gains over uncoded systems. The joint design of the modulation format, labeling rule, and FEC codes is called a coded modulation (CM) scheme. The most widely used CM scheme is Gray-labeled QAM with bit-interleaved coded modulation (BICM), and serves as a benchmark for other CM schemes. In \cite{frey20}, a multilevel coded modulation (MLCM) scheme with SD FEC codes was proposed for the Hurwitz constellation, in which constellation points are a finite set of lattice points from the 4D checkerboard lattice $D_4$, and the boundary is hypercubic. The performance gains over QAM comes from the coding gain of $D_4$. In \cite{stern20, stern21}, CM schemes with non-binary SD codes are designed for the 4D Welti constellation, which has constellation points from the $D_4$ lattice and uses a hypersphere boundary. The performance gains over QAM with BICM comes from the shaping and coding gains of the Welti constellation itself, and the FEC codes (nonbinary codes or multilevel codes) as well. 

However, a CM scheme to preserve VCs' high shaping gains and coding gains after FEC decoding is still lacking. Designing CM schemes for MD VCs that outperforms QAM with BICM is challenging, due to that no Gray labeling exists for MD VCs, and the resulting penalty from a non-Gray labeling might cancel out the shaping and coding gains of VCs. 

In this paper, we focus on MD VCs with a cubic coding lattice, i.e., VCs having high shaping gains but no coding gain. The absent coding gain is instead achieved by FEC codes. We design several CM schemes with SD FEC codes for VCs for the first time, including BICM and MLCM. The considered VCs are of up to $24$ dimensions and have up to $5\times10^{27}$ constellation points with high spectral efficiencies, in order to achieve high shaping gains. However, the proposed labeling rule and log-likelihood ratio (LLR) calculation algorithm have a very low complexity. Moreover, the FEC overhead is lower than commonly used overheads ($15\%$--$25\%$) of high-performance SD FEC codes for optical communications. Thus, the application scenario of the proposed CM scheme would be ultra high-rate transmission systems, such as the upcoming 800 Gbps and 1.25 Tbps standards for fiber communications.





\emph{Notation:} Bold lowercase symbols denote row vectors and bold uppercase symbols denote random vectors or matrices. All-zero and all-one vectors are denoted by $\bzero$ and $\bone$, respectively. Vector inequalities are performed element-wise, e.g., for vectors $\bx,\by\in\R^n$, the inequality $\bx\leq\by$ refers to $x_i\leq y_i$ for $i=1,\ldots,n$. The sets of integer, positive integer, real, complex, and natural numbers are denoted by $\Z$, $\Z^+$, $\R$, $\mathbb{C}$, and $\mathbb{N}$, respectively. Other sets are denoted by calligraphic symbols. Rounding a vector to its nearest integer vector is denoted by $\lfloor \cdot \rceil$, in which ties are broken arbitrarily. The cardinality of a set or the order of a lattice partition is denoted by $|\cdot|$.

\section{Lattices and VCs}
An $n$-dimensional lattice $\Lambda$ is an infinite set of points spanned by the rows of its $n\times n$ generator matrix $\bG_{\Lambda}$ with all integer coefficients, i.e.,
\begin{align}
\Lambda \triangleq \{ \bu \bG_{\Lambda} :\; \bu \in \Z^n \}
.\label{eq:lattice_defination}\end{align}
The \emph{closest lattice point quantizer} of a lattice $\Lambda$, denoted by $\Q_{\Lambda}(\cdot)$, finds the closest lattice point in $\Lambda$ of an arbitrary point $\bx\in\R^n$, i.e.,
\begin{align}
    \Q_{\Lambda}(\bx)=\argmin_{\blambda \in \Lambda}\|\bx-\blambda\|^2.
\end{align}

A sublattice $\Lambda'$ of $\Lambda$, denoted by $\Lambda' \subseteq \Lambda$, contains a subset of the lattice points\footnote{Arbitrary points in $\R^n$ are referred to ``points'' in this paper. To avoid ambiguity, ``lattice point'' is used when a point also belongs to a lattice. Later throughout the paper, ``constellation points'' refers to the points in VCs.} of $\Lambda$, which is spanned by the generator matrix $\bG_{\Lambda'}$ satisfying
\begin{align}
\bG_{\Lambda'}=\bJ\bG_{\Lambda},
\end{align}
with an $ \bJ \in \Z^{n\times n}$. The \emph{lattice partition} $
\Lambda/\Lambda'$ partitions $\Lambda$ into $|\Lambda/\Lambda'|=|\!\det\bG_{\Lambda'}|/|\!\det\bG_{\Lambda}|=|\!\det\bJ|$ disjoint \emph{cosets} of $\Lambda'$ \cite{calderbank87}, and $|\Lambda/\Lambda'|$ is called the \emph{partition order}. If one arbitrary lattice point is selected from each of these cosets, a set of \emph{coset representatives} (not unique) is formed, denoted by $[\Lambda/\Lambda']$. Then every lattice point $\blambda \in\Lambda$ can be written as 
\begin{align}
    \blambda=\bc+\blambda',
\end{align}
where $\blambda' \in \Lambda'$ and $\bc\in[\Lambda/\Lambda']$ can be uniquely labeled by $k=\log_2(|\Lambda/\Lambda'|)$ bits if $|\Lambda/\Lambda'|$ is a power of $2$. The whole lattice $\Lambda$ is decomposed as
\begin{align}
    \Lambda=[\Lambda/\Lambda']+\Lambda'.
\end{align}

A \emph{partition chain}, formed by a sequence of lattices ${\Lambda^{0}\supseteq\Lambda^{1}\supseteq\dots\supseteq\Lambda^{q}}$ with $q\in\Z^+$, is denoted by $\Lambda^{0}/\Lambda^{1}/\dots/\Lambda^{q}$ \cite{forney89a}. Every lattice point $\blambda_0\in\Lambda^0$ can be written as 
\begin{align}
    \blambda_0=\sum_{i=1}^q\bc_i+\blambda_q,
\end{align}
where $\bc_i \in [\Lambda^{i-1}/\Lambda^i]$ for $i=1,\ldots,q$ and $\blambda_q \in \Lambda^q$. If the partition orders $|\Lambda^{i-1}/\Lambda^{i}|$ for $i=1,\ldots,q$ are powers of 2, $\blambda_0$ can be uniquely labeled by the binary tuple
\begin{align}
    \bb=(\bb_1,\bb_2,\ldots,\bb_q),
\end{align}
where $\bb_i$ is the bit labels of $\bc_i$ with the length of $k_i$ for $i=1,\ldots,q$ and the total length of $\bb$ is ${\sum_{i=1}^{q}k_i=\log_2(|\Lambda^0/\Lambda^q|)}$. The lattice $\Lambda^0$ can be decomposed as
\begin{align}
    \Lambda^0=[\Lambda^0/\Lambda^1]+\dots+[\Lambda^{q-1}/\Lambda^q]+\Lambda^q.
\end{align}

An $n$-dimensional VC is a set of coset representatives of a lattice partition $\Lambda_{\text{c}}/\Lambda_{\text{s}}$, where $\Lambdac$ is called the \emph{coding lattice}, $\Lambdas$ is called the \emph{shaping lattice}, and the partition order is $M=|\Lambdac/\Lambdas|$, which is a power of 2 to enable binary labeling with $m=\log_2(M)$ bits. The VC points $\Gamma$ are defined as all lattice points in the translated $\Lambdac$ having the all-zero point $\bzero$ as their closest lattice point in $\Lambdas$, i.e. \cite{forney89b},
\begin{align}
\Gamma \triangleq \{\bx \in (\Lambdac -\ba) :\; \Q_{\Lambdas}(\bx)=\bzero \},
\label{eq:VC}
\end{align}
where the \emph{offset vector} $\ba\in \R^n$ is usually optimized to minimize the average symbol energy \cite{conway83} 
\begin{align}
E_{\text{s}}=\frac{1}{M}\sum_{\bx\in\Gamma}\|\bx\|^{2}.\label{eq:Es}
\end{align}
The \emph{spectral efficiency} \cite{forney89a, kschischang93, agrell09} in bits per two-dimensional (2D) symbol for the uncoded system is defined as 
\begin{align}
    \beta=\frac{2m}{n}~\text{[bits/2D-symbol]}.\label{eq:beta}
\end{align}
The signal-to-noise ratio (SNR) is defined as $E_{\text{s}}/\sigma_{\text{tot}}^2$, where $\sigma_{\text{tot}}^2$ is the total noise variance.



\section{Labeling of VCs}\label{sec:labeling}
\subsection{Encoding and decoding}\label{sec:encdec}
A \emph{labeling} function is a map from binary labels of length $m$ to constellation points. For the considered VCs based on the lattice partition $\Z^n/\Lambdas$, we divide the labeling algorithm into two steps: first from binary labels to integers and then from integers to VC points, i.e.,
$\{0,1\}^m \rightarrow \U \rightarrow \Gamma$. The integer set $\U$ is formed by first writing the generator matrix of the shaping lattice $\Lambdas$ as a lower-triangular form $\bGs$ with diagonal elements $\bh=(h_1,\dots,h_n)$, and then letting 
\begin{align}
    \U =\{\bu:\bzero\leq \bu\leq \bh-\bone\}.
\end{align}

\emph{Encoding:}
The function that maps binary labels to integers is denoted by $f: \; \{0,1\}^m \rightarrow \U$, which will be discussed in Sections~\ref{sec:pseudo},~\ref{sec:setpart}, and~\ref{sec:hybrid}. The algorithm that maps integers $\U$ to constellation points $\Gamma$ was proposed in \cite{kurkoski18} and summarized in \cite[Alg.~1]{ourTC}, which is denoted by ${g: \; \U \rightarrow \Gamma}$ in this paper.

\emph{Decoding:} After receiving a noisy version of a VC point, the algorithm that maps it back to an estimate of the transmitted VC point was proposed in \cite{kurkoski18} and summarized in \cite[Alg.~2]{ourTC}, which is denoted by a function ${w: \; \R^n \rightarrow \U}$ in this paper. Then the binary labels are obtained by the inverse of $f$, i.e., $f^{-1}: \; \U \rightarrow \{0,1\}^m$.


The rest of this section introduces three different mapping functions $f$ for the considered VCs based on the lattice partition $\Z^n/\Lambdas$. Section~\ref{sec:pseudo} reviews the Gray mapping proposed in 
 \cite{ourISIT}. Section~\ref{sec:setpart} proposes a new mapping function, the set partitioning (SP) mapping based on Ungerboeck's SP and lattice partition chains. Another new hybrid mapping function combining the SP mapping and pseudo-Gray mapping is then proposed in Section~\ref{sec:hybrid}.

\subsection{Gray mapping}\label{sec:pseudo}
In \cite{ourISIT}, a mapping method between binary labels and integers is proposed in order to minimize the uncoded BER of VCs, which works in the following way.

First, the binary label $\bb\in \{0,1\}^m$ is divided into $n$ blocks according to $\bh$,
\begin{align}
    \bb=(\bb_1,\bb_2,\ldots,\bb_n),\notag
\end{align}
each of which has $\log_2(h_i)$ bits for $i=1,\ldots,n$, and $\sum_{i=1}^n{\log_2(h_i)}=m$. Then $\bb_i$ is converted to an integer $u_i$ using the binary reflected Gray code (BRGC) \cite{agrellBRGC04} for $i=1,\ldots,n$, yielding 
\begin{align}
    \bu=(u_1,\ldots,u_n).
\end{align}
The above procedures converting $\bb$ to $\bu$ according to the BRGC is denoted by $f_{\text{BRGC}}(\bb,\bh)$, and the inverse process of converting an integer vector $\bu$ to a binary vector $\bb$ is denoted by $f^{-1}_{\text{BRGC}}(\bu,\bh)$ in this paper. After mapping integers to VC points using function $g$ defined in section~\ref{sec:encdec}, the labeling is not true Gray, but close to Gray, which is called ``pseudo-Gray'' labeling.

\subsection{SP mapping}\label{sec:setpart}

\begin{figure}[tbp]
    \centering
    \includegraphics[width=2.6in]{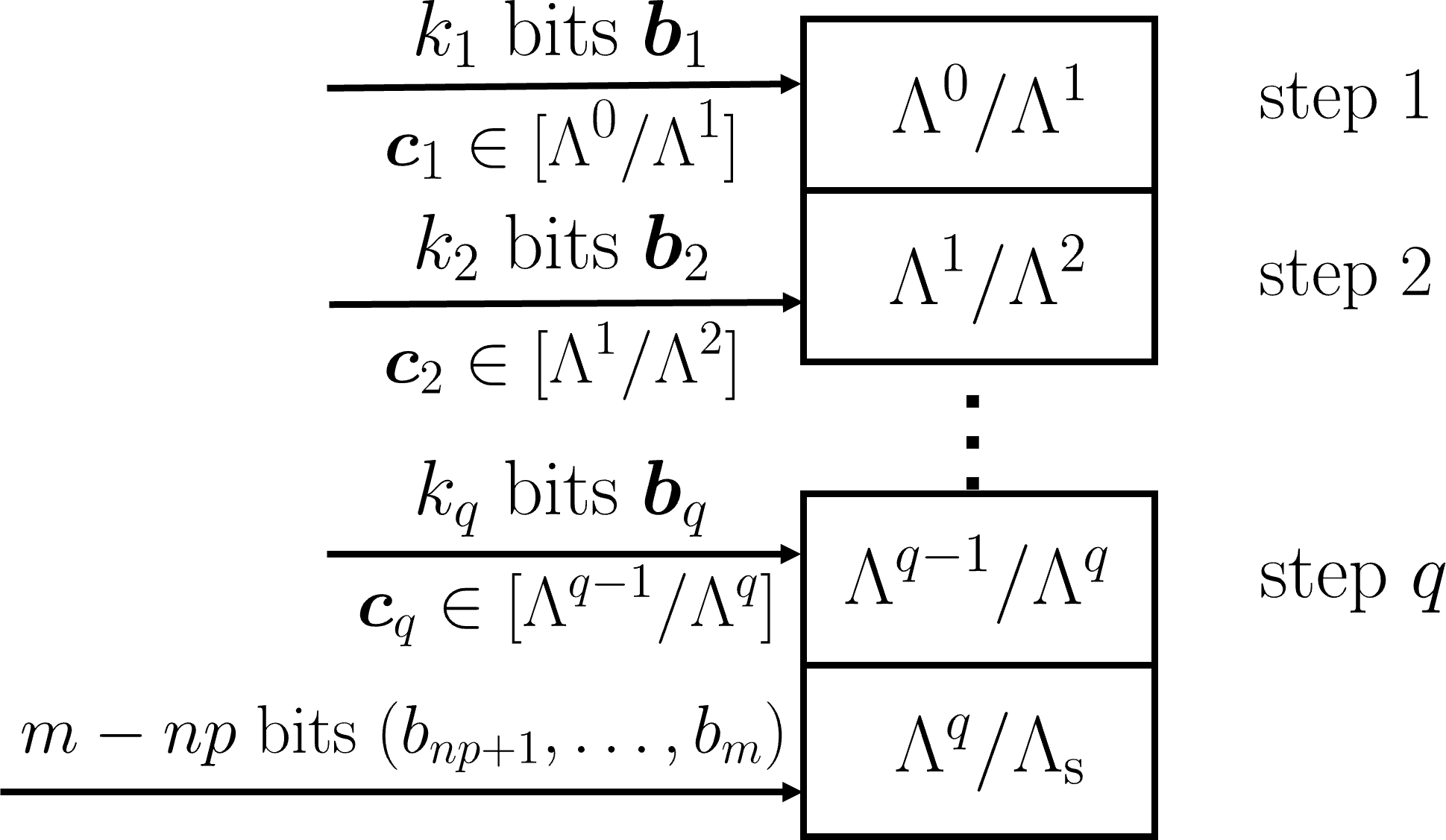}
    \caption{Illustration of the labeling of a partition chain $\Lambda^0/\Lambda^1/\dots/\Lambda^{q-1}/\Lambda^q/\Lambdas$.}
    \label{fig:partition chain}
\end{figure}

\begin{table}[tbp]
  \renewcommand{\arraystretch}{1.2}
  \renewcommand{\tabcolsep}{12pt}
  \caption{Example partition chains in the SP mapping for MD VCs with a cubic coding lattice $\Z^n$.}
  \label{tab:partition chain}
  \centering
  \begin{tabular}{l l l l l l}
    \hline 
    $n=2$ & \multicolumn{5}{c}{$\Z^2/D_2/2\Z^2/2D_2/4\Z^2/4D_2/\dots$} \\
    \hline 
    Step $i$ & 1 & 2 &3 &4 &5\\
    $k_i$ & 1 & 1& 1 &1&1\\
    $d_i^2$ & 2 &4& 8&16&32\\
    \hline \hline
    $n=4$ & \multicolumn{5}{c}{$\Z^4/D_4/2\Z^4/2D_4/4\Z^4/4D_4/\dots$} \\
    \hline 
    Step $i$ & 1 & 2 &3 &4 &5\\
    $k_i$ & 1 & 1& 1 &1&1\\
    $d_i^2$ & 2 &4& 8&16&32\\
    \hline \hline
    $n=8$ & \multicolumn{5}{c}{$\Z^8/D_8/E_8\bR_{8}/2E_8/2E_8\bR_{8}/4E_8/\dots$} \\
    \hline 
    Step $i$ & 1 & 2 &3 &4 &5\\
    $k_i$ & 1 & 3& 4 &4&4\\
    $d_i^2$ & 2 & 4& 8 &16& 32\\
    \hline \hline
        $n=16$ & \multicolumn{5}{c}{$\Z^{16}/D_{16}/D_{16}\bR_{16}/\Lambda_{16}/\Lambda_{16}\bR_{16}/2\Lambda_{16}/\dots$} \\
    \hline 
    Step $i$ & 1 & 2 &3 &4 &5\\
    $k_i$ & 1 & 8 & 3 & 8 &8\\
    $d_i^2$ & 2 & 4& 8 &16& 32\\
    \hline
  \end{tabular}
\end{table}

\begin{table}[tbp]
  \renewcommand{\arraystretch}{1.2}
  \renewcommand{\tabcolsep}{12pt}
  \caption{An example look-up table for the coset representatives of the lattice partition $D_8/E_8\bR_8$ and their bit labels.}
  \label{tab:CosetRep}
  \centering
  \begin{tabular}{l l}
    \hline
    $[D_8/E_8\bR_8]$ & labels \\
    \hline \hline
    $(0 0 0 0 0 0 0 0)$ & $(0 0 0)$\\
    $(0 1 0 1 0 0 0 0)$ & $(0 0 1)$ \\
    $(0 0 0 1 1 0 0 0)$ & $(0 1 0)$ \\
    $(0 1 0 0 1 0 0 0)$ & $(0 1 1)$ \\
    $(1 1 0 0 0 0 0 0)$ & $(1 0 0)$ \\
    $(1 0 0 1 0 0 0 0)$ & $(1 0 1)$ \\
    $(1 1 0 1 1 0 0 0)$ & $(1 1 0)$ \\
    $(1 0 0 0 1 0 0 0)$ & $(1 1 1)$ \\
    \hline
  \end{tabular}
\end{table}
Ungerboeck's SP concept \cite{ungerboeck82} maps binary labels to 1D or 2D constellation points by successively partitioning the constellation into two subsets at each bit level in order to maximize the intra-set minimum squared Euclidean distance (MSED) at each level, so that unequal error protection can be implemented on different bit levels. Since all partition orders are 2, Ungerboeck's SP is also called binary SP. Binary SP has been applied to 1D, 2D \cite{wachsmann99,isaka98,yuan6g21}, and 4D \cite{beygi14jlt,frey20,stern20,stern21} signal constellations. When binary SP is applied in larger than 2 dimensions, the MSED might not increase at every bit level. Binary SP requires one encoder and one decoder at each bit level, which has a high complexity in FEC for large constellations. Generalized from the binary SP, signal sets can be partitioned into multiple subsets based on the concept of cosets \cite{calderbank87,forney88,forney89b,wei87}, which enables SP in higher dimensions \cite{wei87,forney88,forney89b} and increasing MSED at every partition level. How the coset representatives are labeled at each partition level is not specified. In this section, we introduce a systematic algorithm for mapping bits $\bb\in\{0,1\}^m$ to integers $\bu\in\U$ based on SP such that the MSED doubles at every partition level for very large MD VCs based on the lattice partition $\Z^n/\Lambdas$.

For large constellations, after getting a sufficiently large intra-set MSED, it is reasonable to not partition the remaining subsets and leave the corresponding bit levels uncoded. One convenient way is to stop partitioning when a scaled integer lattice $2^p\Z^n$ is obtained, where $p\in \mathbb{N}$. Then we map the last $m-np$ bits to integers according to BRGC to minimize the BER for the uncoded bits.

The preprocessing of the proposed SP mapping works as follows. First, $q$ intermediate lattices $\Lambda^1, \dots, \Lambda^{q}$ are found to form the partition chain
\begin{align} \Lambda^0/\Lambda^1/\dots/\Lambda^{q}/\Lambdas,
\end{align}
where $\Lambda^0=\Z^n$ and $\Lambda^q=2^p\Z^n$ is where to stop the partition. The partition chain should satisfy $\Lambda^0 \supset \Lambda^1 \supset \dots \supset \Lambda^q \supseteq \Lambdas$ and have increasing MSEDs of $d_i^2=2^i$ for ${i=1,\ldots,q}$. The order of each partition step $|\Lambda^{i-1}/\Lambda^i|=2^{k_i}$ for $i=1,\ldots,q$ and $\sum_{i=1}^q k_i= \log_2(|\Lambda^0/\Lambda^q|)=np$. At every partition step $i$, all coset representatives $[\Lambda^{i-1}/\Lambda^i]$ are labeled by $k_i$ bits and the mapping rules are stored in a look-up table $\bC_i$. Conventionally, the set of coset representatives contains the all-zero lattice point labeled by the all-zero binary tuple. Fig.~\ref{fig:partition chain} illustrates the mapping for the partition chain in general. Table~\ref{tab:partition chain} lists some example partition chains and their intra-set MSEDs for MD VCs with a cubic coding lattice. These partition chains contain commonly used lattices as intermediate lattices including the $n$-dimensional checkerboard lattice $D_n$, $8$-dimensional (8D) Gosset lattice $E_{8}$, $16$-dimensional (16D) Barnes--Wall lattice $\Lambda_{16}$ \cite{pook23}, and the $24$-dimensional (24D) Leech lattice $\Lambda_{24}$ \cite[Ch.~4]{conway99book}. The $n\times n$ matrix $\bR_n$ is an integer orthonormal rotation matrix with a determinant of $\det\bR_n=2^{n/2}$ \cite{forney89b, forney88}. When multiplied with a lattice generator matrix on the right, it rotates every two dimensions of the lattice by $\ang{45}$ and rescales it by $\sqrt{2}$. The size of the look-up table $\bC_i$ is $2^{k_i}$, which is not more than $2^8$ in Table~\ref{tab:partition chain} and much smaller than a table for the whole VC.

As an example, a set of coset representatives of the partition $D_8/E_8\bR_8$ and one set of possible bit labels are listed in Table~\ref{tab:CosetRep}. Note that neither the choice of the set of coset representatives nor the mapping within the look-up table is unique, and both of them are arbitrarily selected. The effects of different choices on the performance are not studied. However, we conjecture that there would be no big difference since the intra-set MSED cannot be increased by further partitioning the set of coset representatives.

The SP demapping $f^{-1}_{\text{SP}}$ from an integer vector $\bu$ to its bit labels $\bb$ works as follows. Given an integer vector $\bu \in \U$, starting from the first partition step $\Lambda^0/\Lambda^1$, we know that $\bu$ belongs to one coset of this partition since $\bu\in \Lambda^0=\Z^n$. By full search among a certain set of coset representatives $[\Lambda^0/\Lambda^1]$,  there must be only one $\bc_1\in[\Lambda^0/\Lambda^1]$ such that $(\bu-\bc_1)\cdot\bG_{\Lambda^1}^{-1}$ yields an integer vector, where $\bG_{\Lambda^1}$ is the generator matrix of $\Lambda^1$. The bit labels $\bb_1$ corresponding to $\bc_1$ is found in the look-up table $\bC_1$. Then $\bu-\bc_1$ is a lattice point of $\Lambda^1$, which must belong to a certain coset of $\Lambda^1/\Lambda^2$. A vector $\bc_2\in[\Lambda^1/\Lambda^2]$ is found such that $(\bu-\bc_1-\bc_2)\cdot \bG_{\Lambda^{2}}^{-1}$ yields an integer vector, where $\bG_{\Lambda^{2}}$ is the generator matrix of $\Lambda^2$ and the bit labels $\bb_2$ corresponding to $\bc_2$ is found in the look-up table $\bC_2$. The procedure is repeated until all $\bc_i$ and $\bb_i$ are obtained for $i=1,\ldots,q$. 

Next, the remaining $m-np$ bit labels for the partition $2^p\Z^n/\Lambdas$ are found as follows. First, the coset representatives of the partition $\Z^n/2^p\Z^n$ are set as 
\begin{align}
    \mathcal{S}=[\Z^n/2^p\Z^n]=\{\bs\in\Z^n:\bzero \leq \bs \leq (2^p-1) \cdot \bone\}.\label{eq:SP_coset}
\end{align} 
There must be a unique $\bs\in\mathcal{S}$ such that $\bu-\bs\in 2^p\Z^n$, and $\bs$ can be easily found by 
\begin{align}
    \bs=\bu \bmod 2^p,
\end{align}
where $\bmod$ is the modulo operator that takes the remainder of a vector element-wise and returns a vector. The purpose of choosing such a set of coset representatives is to make sure that $\bu-\bs$ still falls within the range of $\U$, i.e., $\bu-\bs \in \U \cap 2^p\Z^n$. Then we know that
\begin{align}
    \frac{\bu-\bs}{2^p} \in \Z^n
\end{align}
with the range
\begin{align}
    \bzero\leq\frac{\bu-\bs}{2^p}\leq \frac{\bh}{2^p}-\bone.
\end{align}
The bit labels of $(\bu-\bs)/2^p$ can be obtained by converting each decimal element to bits according to BRGC, i.e.,
\begin{align}\label{eq:leftbitsBRGC}
(b_{np+1},\dots,b_m)=f^{-1}_{\text{BRGC}}\left(\frac{\bu-\bs}{2^p},\frac{\bh}{2^p}\right).
\end{align}

Now we describe the SP mapping $f_{\text{SP}}$ from the bit labels $\bb$ to the integer vector $\bu$. Given bit labels $\bb\in\{0,1\}^m$, the first $np$ bits are divided into $q$ blocks $\bb_i$ for $i=1,\ldots,q$, each of which has $k_i$ bits and indicates a coset representative $\bc_i$ according to the look-up $\bC_i$. Then $\bc=\sum_{i=1}^q \bc_i$ indicates a coset representative of the lattice partition $\Z^n/2^p\Z^n$, but $\bc$ might not belong to $\mathcal{S}$, which can be converted to $\bs\in \mathcal{S}$ by
\begin{align}
    \bs=\bc \bmod 2^p.
\end{align}
Now we know that $\bu-\bs\in2^p\Z^n$. The remaining $m-np$ bits of $\bb$ indicate an integer vector
\begin{align}\label{eq:t}
 \bt= f_{\text{BRGC}}\left((b_{np+1},\dots,b_m),\frac{\bh}{2^p}\right).
\end{align}
Finally, $\bu$ is obtained by
\begin{align}\label{eq:u}
    \bu=\bs+2^p\bt.
\end{align}

Algorithms~\ref{Alg:map1} and~\ref{Alg:demap1} summarize the SP mapping process between $\bu$ and $\bb$ for VCs with a cubic coding lattice. 

\begin{algorithm}[tbp]
	\caption{SP mapping $f_\text{SP}$}\label{Alg:map1} 
	Input: $\bb$. Output: $\bu$.\\
	Preprocessing: The partition chain $\Lambda^0/\Lambda^1/\dots/\Lambda^q/\Lambdas$ is given, where all partition orders $|\Lambda^{i-1}/\Lambda^i|$ for $i=1,\ldots,q$ are powers of 2. Set the look-up tables $\bC_i$ between all coset representatives and their corresponding bit labels for all partition steps. Divide the first $np$ bits of $\bb$ into $q$ blocks $\bb_i$ for $i=1,\ldots,q$, each of which has $k_i$ bits. Find a lower-triangular generator matrix $\bGs$ of $\Lambdas$ and denote the diagonal elements of $\bGs$ as $\bh$.
	\begin{algorithmic}[1]
        \State Find the corresponding $\bc_i$ of $\bb_i$ according to $\bC_i$ for ${i=1,\ldots,q}$.
	    \State Let $\bc \leftarrow\sum_{i=1}^{q}\bc_i$
            \State Let $\bs \leftarrow \bc- \lfloor\bc/2^p\rfloor \cdot 2^p$
	    \State Let $\bt\leftarrow f_{\text{BRGC}}((b_{np+1},\dots,b_m),\bh/2^p)$
	    \State Let $\bu \leftarrow \bs+2^p\bt$
	\end{algorithmic} 
\end{algorithm}

\begin{algorithm}[tbp]
	\caption{SP demapping $f^{-1}_{\text{SP}}$} \label{Alg:demap1} 
	Input: $\bu$. Output: $\bb$.\\
        The partition chain $\Lambda^0/\Lambda^1/\dots/\Lambda^q/\Lambdas$ is given, where all partition orders $|\Lambda^{i-1}/\Lambda^i|$ are powers of 2. Set the look-up tables $\bC_i$ between all coset representatives and their corresponding bit labels for all steps $i=1,\ldots,q$. Set the set of coset representatives of the partition $\Lambda^0/\Lambda^q$ as $\mathcal{S}$ defined in \eqref{eq:SP_coset}. Find a lower-triangular generator matrix $\bGs$ of $\Lambdas$ and denote the diagonal elements of $\bGs$ as $\bh$.

	\begin{algorithmic}[1]
         \State Let $\bv\leftarrow \bu$
	    \For {$i=1,\ldots,q$}
        \State Find the only $\bc_i \in [\Lambda^{i-1}/\Lambda^i]$ such that $\bu-\bc_i\in\Lambda^i$
        \State Let $\bb_i$ be the bit labels of $\bc_i$ according to $\bC_i$
	\State Let $\bu \leftarrow \bu - \bc_i $
	    \EndFor
        \State Let $\bs\leftarrow \bv \bmod 2^p$
        \State Let $(b_{np+1},\dots,b_m)=f^{-1}_{\text{BRGC}}\left((\bv-\bs)/2^p,\bh/2^p\right)$
        \State Let $\bb\leftarrow(\bb_1,\dots,\bb_{q},b_{np+1},\dots,b_m)$

	\end{algorithmic} 
\end{algorithm}

\subsection{Hybrid mapping}\label{sec:hybrid}

This mapping is a special case of the SP mapping, which is carefully designed for the considered VCs based on the lattice partition $\Z^n/\Lambdas$. The idea is to only consider $q$ intermediate lattices which are a multiple of the cubic lattice, i.e., $\Lambda^i=2^{p_i}\Z^n$ for $i=0,\ldots,q$ with positive integers ${p_1<p_2<\ldots<p_q=p}$ and $p_0=0$. This yields a partition chain $2^{p_0}\Z^n/2^{p_1}\Z^n/2^{p_2}\Z^n/\dots/2^{p_q}\Z^n/\Lambdas$. Thus, the order of each partition step is $|\Lambda^{i-1}/\Lambda^i|=2^{k_i}=2^{n(p_i-p_{i-1})}$ for $i=1,\ldots,q$ and $\sum_{i=1}^q k_i= \log_2(|\Lambda^0/\Lambda^q|)=np$. The intra-set MSED is $d_i^2=2^{p_i}$ at the $i$th partition step. The coset representatives in each partition step is simply set as  
\begin{align}
    \bC_i=[2^{p_{i-1}}\Z^n/2^{p_i}\Z^n]=\{\bc: \bzero \leq \bc \leq (2^{p_{i}-p_{i-1}}-1)\cdot\bone\},\label{eq:coset_reps}
\end{align}
for $i=1,\ldots,q$, which is labeled by $k_i=n(p_i-p_{i-1})$ bits for $i=1,\ldots,q$. Thanks to that these intermediate lattices are a multiple of $\Z^n$, no full search from $\bC_i$ is needed to find the unique coset representative $\bc_i \in \bC_i$ as in the SP mapping. The coset representative $\bc_i$ can be mapped to $n(p_i-p_{i-1})$ bits by directly converting each element of $\bc_i$ to $p_i-p_{i-1}$ bits according to BRGC, i.e.,
\begin{align}
    \bb_i=f^{-1}_{\text{BRGC}}(\bc_i,2^{p_i-p_{i-1}}\cdot\bone),\label{eq:c2b}
\end{align}
for $i=1,\ldots,q$. 

The hybrid demapping $f^{-1}_{\text{H}}$ from an integer vector $\bu$ to its bit labels $\bb$ works as follows. Given an integer vector $\bu \in \U$, starting from the first partition step $\Z^n/2^{p_1}\Z^n$, the unique $\bc_1\in\bC_1$ can be easily found by 
\begin{align}
    \bc_1=\bu \bmod 2^{p_1}.
\end{align}
 Then $\bu-\bc_1$ is a lattice point of $2^{p_1}\Z^n$, which must belong to a certain coset of $2^{p_1}\Z^n/2^{p_2}\Z^n$. Then the unique $\bc_2$ is obtained by
 \begin{align}
     \bc_2=(\bu-\bc_1) \bmod 2^{p_2}.
 \end{align}
 Repeating this procedure, $\bc_i$ is found successively by
 \begin{align}\label{eq:ciH}
     \bc_i=\biggl(\bu-\sum_{j=1}^{i-1}\bc_j\biggr) \bmod 2^{p_i}
 \end{align}
 for $i=1,\ldots, q$. Then the first $np$ bit labels of $\bu$ can be obtained by \eqref{eq:c2b}. Similar to the SP mapping, the remaining $m-np$ bits are obtained by \eqref{eq:leftbitsBRGC}, where $\bs=\sum_{i=1}^p \bc_i$ in this case.


The hybrid mapping $f_{\text{H}}$ finding the corresponding integer vector $\bu$ of bit labels $\bb$ works as follows. Given bit labels $\bb\in\{0,1\}^m$, $\bc_i \in \bC_i$ can be directly obtained by
\begin{align}
    \bc_i =f_{\text{BRGC}}(\bb_i,2^{p_i-p_{i-1}}\cdot\bone).\label{eq:b2c}
\end{align}
The integer vector $\bs=\sum_{i=1}^p \bc_i$ must belong to the set $\mathcal{S}$ defined in \eqref{eq:SP_coset} due to the definition of $\bC_i$ in \eqref{eq:coset_reps}. Then $\bu$ is obtained combining \eqref{eq:u} and \eqref{eq:t}. 

Algorithms~\ref{Alg:map2} and~\ref{Alg:demap2} summarize the hybrid mapping process between $\bu$ and $\bb$ for VCs with a cubic coding lattice. 
\begin{algorithm}[tbp]
	\caption{Hybrid mapping $f_{\text{H}}$}\label{Alg:map2} 
	Input: $\bb$. Output: $\bu$.\\
       Preprocessing: Given the partition chain $2^{p_0}\Z^n/2^{p_1}\Z^n/2^{p_2}\Z^n/\dots/2^{p_q}\Z^n/\Lambdas$ with positive integers ${p_1<p_2<\ldots<p_q=p}$ and $p_0=0$, set $q$ sets of coset representatives $\bC_i$ as in \eqref{eq:coset_reps} for $i=1,\ldots,q$. Divide the first $np$ bits of $\bb$ into $q$ blocks $\bb_i$ for $i=1,\ldots,q$, each of which has $n(p_i-p_{i-1})$ bits. Find a lower-triangular generator matrix $\bGs$ and denote the diagonal elements of $\bGs$ as $\bh$.
	\begin{algorithmic}[1]
	    \State Let $\bc_i \leftarrow f_{\text{BRGC}}(\bb_i,2^{p_i-p_{i-1}}\cdot\bone)$ for $i=1,\ldots,q$
     \State Let $\bs \leftarrow \sum_{i=1}^q\bc_i$
	    \State Let $\bt \leftarrow f_{\text{BRGC}}\left((b_{np+1},\ldots,b_{m}),\bh/2^{p}\right)$
	    \State Let $\bu \leftarrow \bs +2^{p}\bt$
	\end{algorithmic} 
\end{algorithm}

\begin{algorithm}[tbp]
	\caption{Hybrid demapping $f^{-1}_{\text{H}}$} \label{Alg:demap2} 
	Input: $\bu$. Output: $\bb$.\\
       Preprocessing: Given the partition chain $2^{p_0}\Z^n/2^{p_1}\Z^n/2^{p_2}\Z^n/\dots/2^{p_q}\Z^n/\Lambdas$ with positive integers ${p_1<p_2<\ldots<p_q=p}$ and $p_0=0$, set $q$ sets of coset representatives $\bC_i$ as in \eqref{eq:coset_reps} for $i=1,\ldots,q$. Find a lower-triangular generator matrix $\bGs$ and denote the diagonal elements of $\bGs$ as $\bh$.
	\begin{algorithmic}[1]
           \For {$i=1,\ldots,q$}
        \State Let $\bc_i=\bu \bmod 2^{p_i}$
        \State Let $\bb_i \leftarrow f^{-1}_{\text{BRGC}}(\bc_i,2^{p_i-p_{i-1}}\cdot\bone)$
	\State Let $\bu \leftarrow \bu - \bc_i $
	    \EndFor
        \State Let $\left(b_{np+1},\ldots,b_{m}\right) \leftarrow f^{-1}_{\text{BRGC}}(\bu/2^{p},\bh/2^{p})$
        \State Let $\bb=(\bb_1,\ldots,\bb_q,b_{np+1},\dots,b_m)$
	\end{algorithmic} 
\end{algorithm}

\begin{figure}[tbp]
\centering
\begin{subfigure}{\linewidth}
    \centering
    \includegraphics[width=3.5in]{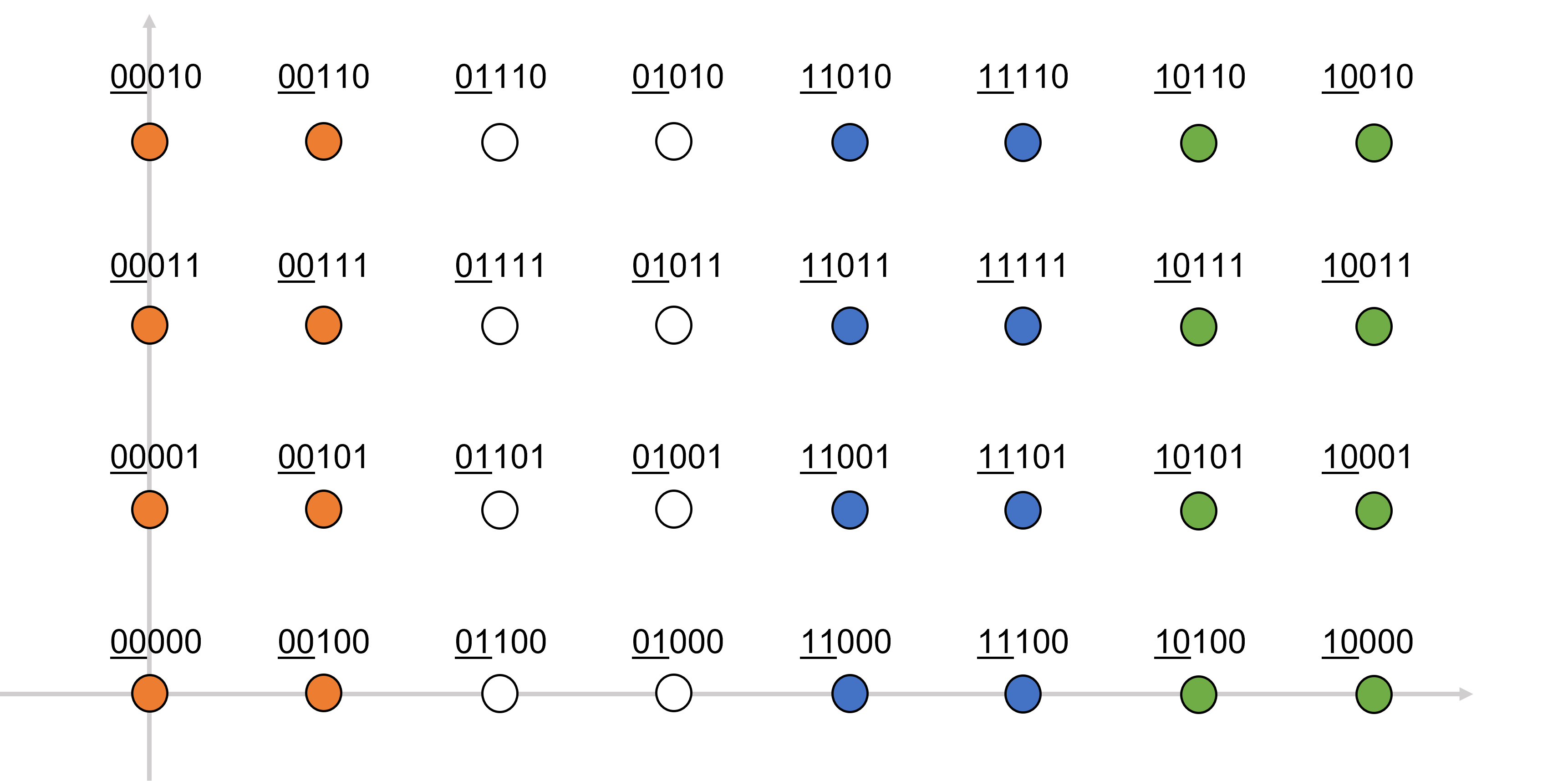}
    \caption{Gray mapping}
    \label{Fig:labeling1}
    \end{subfigure}
    \begin{subfigure}{\linewidth}
        \centering
        \includegraphics[width=3.5in]{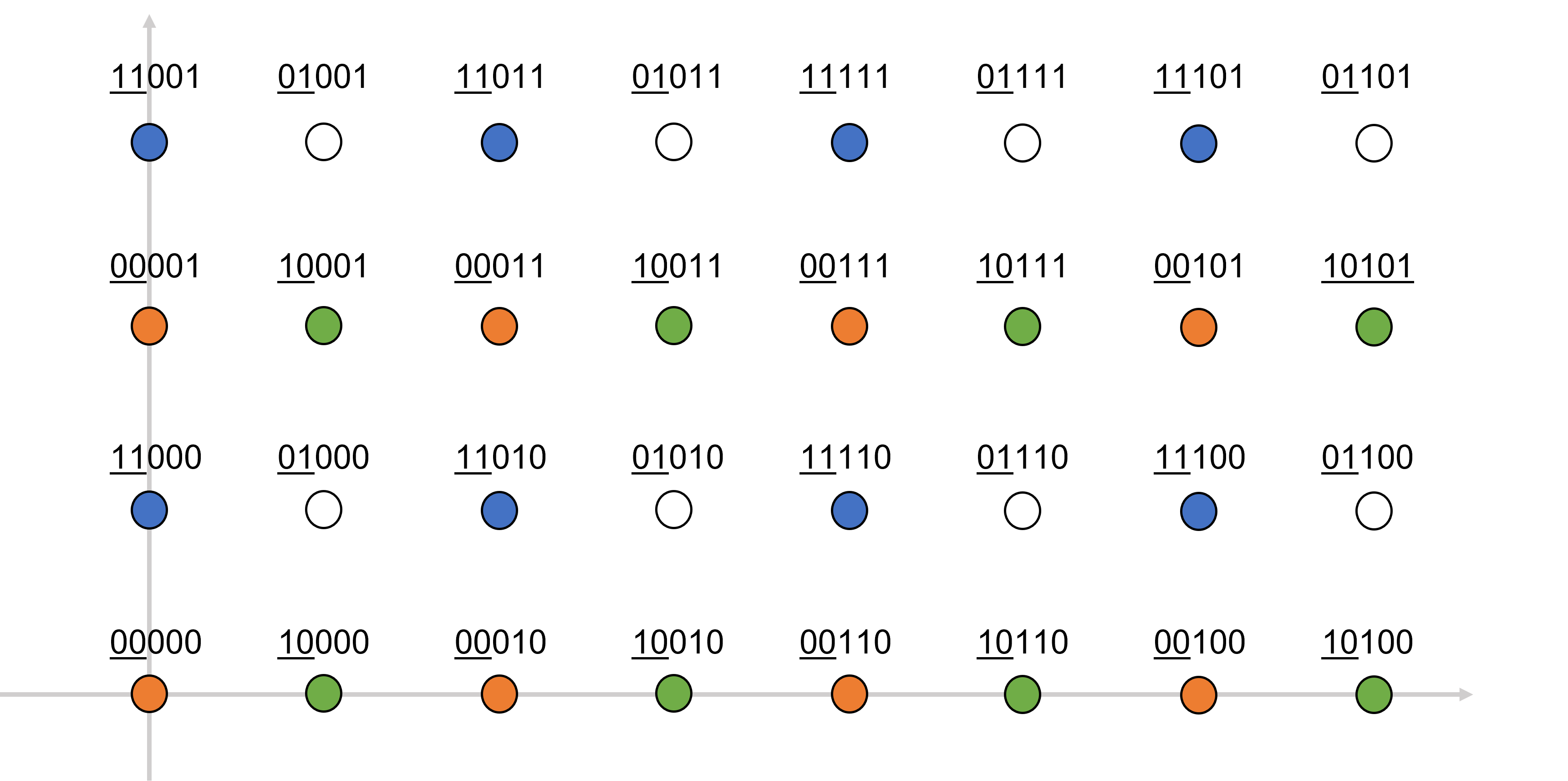}%
        \caption{SP mapping}
        \label{Fig:labeling2}
    \end{subfigure}
    \begin{subfigure}{\linewidth}
        \centering
        \includegraphics[width=3.5in]{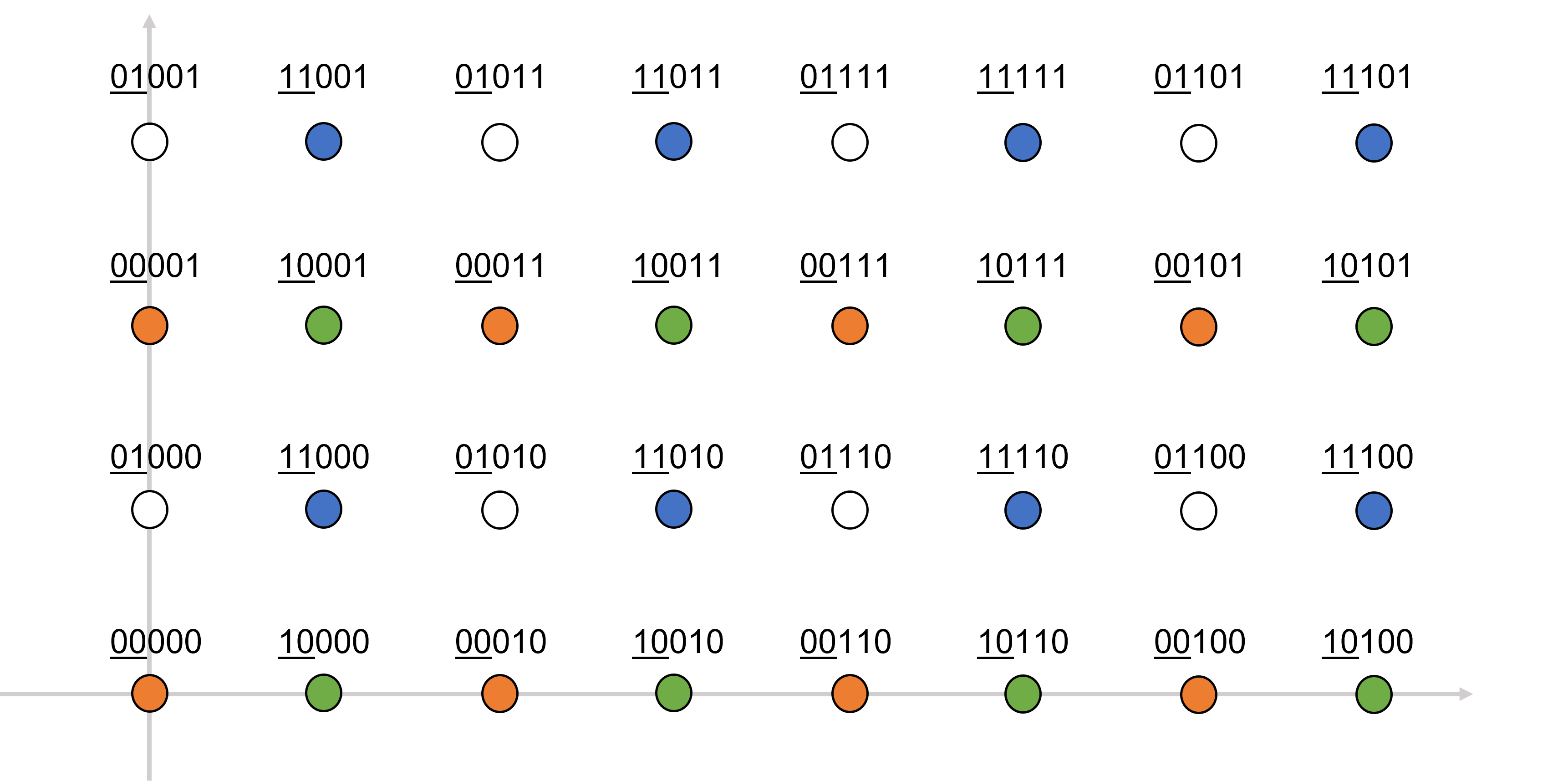}%
        \caption{Hybrid mapping}
        \label{Fig:labeling3}
    \end{subfigure}
    \caption{Example 1: Different mapping rules $f$ between integer vectors and bit labels for the VC based on the lattice partition $\Z^2/4D_2$. Integer points having the same first two bits are filled with the same color for better visualization and comparison.}
    \label{Fig:labeling}
\end{figure}

\emph{Example 1:} A simple example is a 2D VC based on the lattice partition $\Z^n/4D_2$, where $4D_2$ is the scaled 2D checkerboard lattice with the generator matrix
\begin{align}
\bGs=\begin{pmatrix}
8 & 0\\ 
4 & 4
\end{pmatrix}.\notag
\end{align}
This VC does not provide any shaping gain, which is just for simplicity of illustration. Fig.~\ref{Fig:labeling} illustrates the three different mapping rules $f$ for this example VC. It can be observed that all integer points have only 1-bit difference from their nearest neighbors in the Gray mapping. The SP mapping is based on the lattice partition chain $\Z^2/D_2/2\Z^2/4D_2$ ( $q=2$, $p=1$). The hybrid mapping is based on the lattice partition chain $\Z^2/2\Z^2/4D_2$ ($q=p=1$), and ${[\Z^n/2\Z^n]=\{(0,0),(1,0),(0,1),(1,1)}\}$. For both SP and hybrid mapping, within each coset of $2\Z^2/4D_2$, the points have only 1-bit difference among the last three bits from their closest neighbors. 

\section{CM schemes}\label{sec:CM}

The joint design of forward error correction (FEC) coding and modulation formats is called coded modulation (CM), which plays a vital part in modern communication systems. Designing a CM scheme involves a trade-off among the spectral efficiency, power, and complexity. In this section, we propose three SD CM schemes for VCs based on the lattice partition $\Z^n/\Lambdas$, adopting the three labeling schemes introduced in section~\ref{sec:labeling}, and the computation of the log-likelihood ratios (LLRs) for SD decoding is discussed.

The designed CM schemes can be combined with an outer hard-decision (HD) code, known as concatenated coding \cite{forney1965concatenated,barakatain2020}. Concatenated codes are widely used in many communication standards nowadays, such as the DVB-S2 standards \cite{dvbs2} for satellite communications and the 400ZR \cite{400ZR} and upcoming 800G standards for fiber-optical communications \cite{stern21}. The inner CM scheme brings the uncoded BER down to a certain target BER (e.g., around $10^{-3}$ for fiber-optic communications). Then the outer code can further eliminate the error floor and achieve a very low BER as needed. Commonly used outer codes include Reed--Solomon codes \cite{wicker1999reed}, turbo product codes \cite{wang23}, Bose--Chaudhuri--Hocquenghem codes \cite{forney1965bch}, staircase codes \cite{smith2011staircase}, and zipper codes \cite{sukmadji2019zipper}.
\begin{figure*}[tbp]
\centering
\begin{subfigure}{\linewidth}
    \centering
    \includegraphics[width=5.4in]{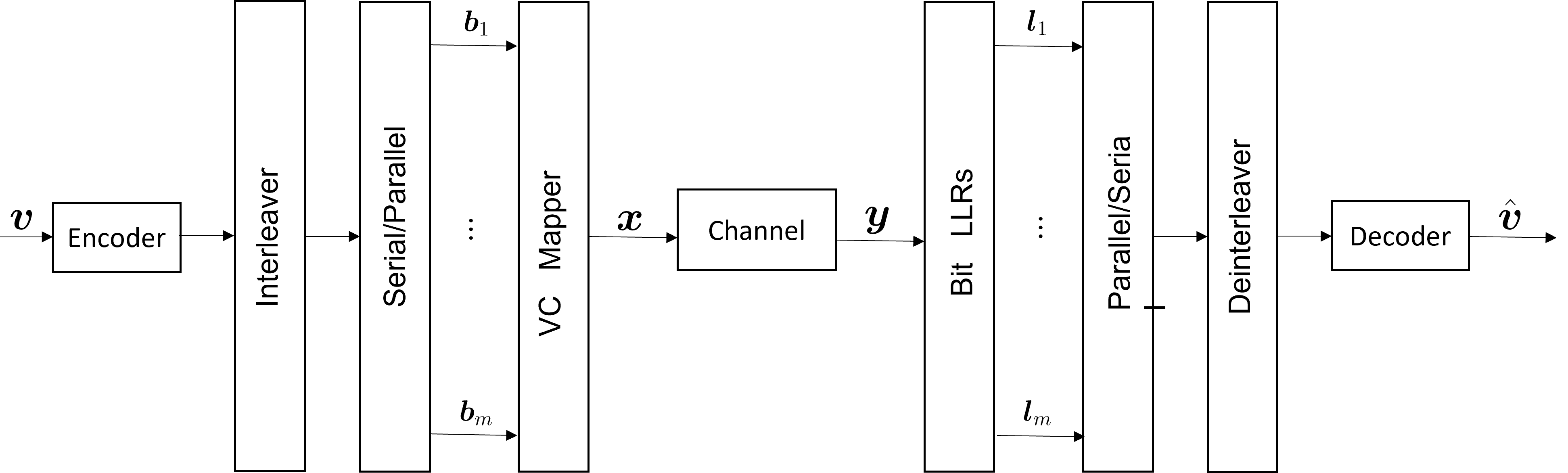}
    \caption{BICM: A block of $N\Rc$ information bits $\bv$ are encoded into $N$ bits by the encoder and then permuted by the interleaver to avoid burst errors \cite{caire98}. The serial bits after interleaving are converted to $m$ parallel bit streams $\bb_1,\ldots,\bb_m$ of length $N/m$. At the time slot $j=1,\ldots,N/m$, a VC mapper first maps $m$ bits $\bb^j=(b_1^j,\dots,b_m^j)$ to an integer $\bu^j$ by $\bu^j=f_{\text{BRGC}}\left(\bb^j,\bh\right)$, and then maps $\bu^j$ to a VC point $\bx^j \in \Gamma$ by $\bx^j=g(\bu^j)$. The receiver deinterleaves $N$ independent LLRs of the bits and then uses them to decode $\hat{\bv}$.}
    \label{Fig:BICM}
    \end{subfigure}
    \begin{subfigure}{\linewidth}
        \centering
        \includegraphics[width=7in]{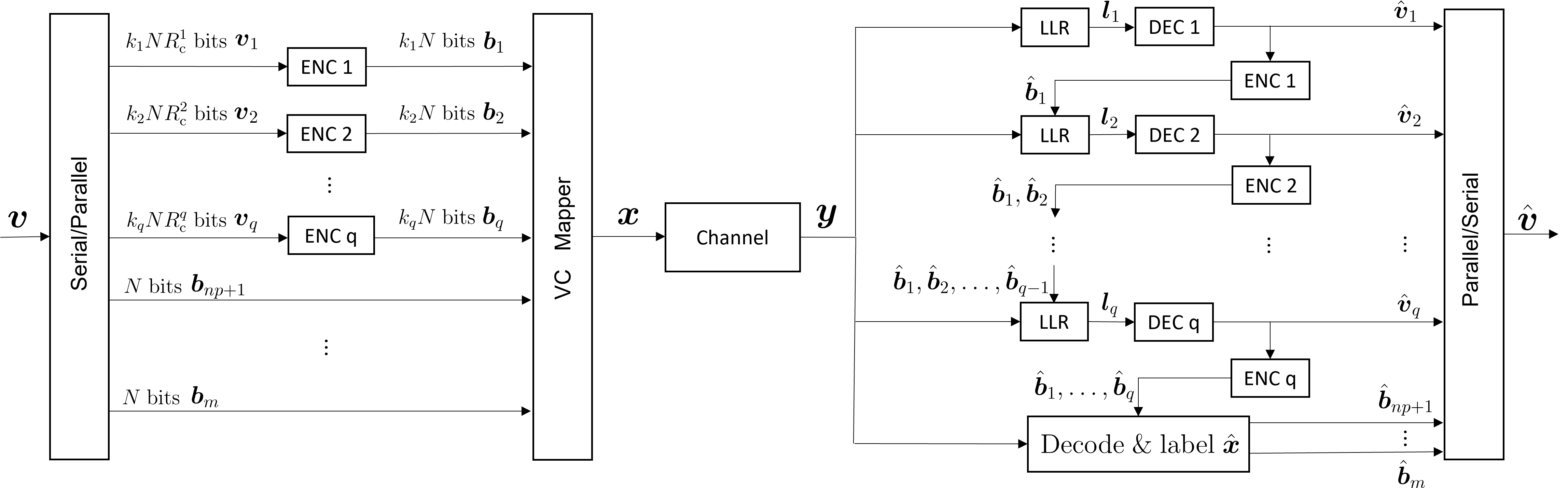}%
        \caption{MLCM: A block of serial information bits $\bv$ are partitioned into $q$ parallel bit streams $\bv_i$ with $k_iN\Rc^{i}$ bits for $i=1,\ldots,q$ and $m-np$ parallel uncoded bit streams $\bb_{np+1},\ldots,\bb_m$ with length $N$. Then $\bv_i$ is encoded into $k_iN$ bits $\bb_i$ by encoder (ENC) $i$ for $i=1,\ldots,q$. The VC mapper first maps bits to $N$ integer vectors by the SP or hybrid mapping, and then encode these integer vectors into $N$ VC points $\bx$. Multistage decoding is performed at the receiver after receiving $N$ noisy symbols $\by$. Decoder (DEC) 1 first decodes $k_1N\Rc^1$ bits $\hat{\bv}_1$ back based on $\by$ and $k_1N$ LLRs $\bl_1$. Then $\hat{\bv}_1$ is encoded into $\hat{\bb}_1$ by encoder 1. Decoder $i=2,\ldots,q$ successively decodes $\hat{\bv}_i$ and reencodes it into $\hat{\bb}_i$ based on $\by$ and LLRs $\bl_i$, given all previous bits $\hat{\bb}_1,\dots,\hat{\bb}_{i-1}$. The estimation of the uncoded bits $\hat{\bb}_{np+1},\ldots,\hat{\bb}_m$ is obtained after getting $\hat{\bb}_1,\dots,\hat{\bb}_{q}$.}
        \label{Fig:MLCM}
    \end{subfigure}
    \caption{Block diagrams of BICM and MLCM for VCs.}
    \label{Fig:CM}
\end{figure*}


\subsection{BICM for VCs with Gray mapping}
Consider a channel with input symbols $\bX$ labeled by $m$ bits $(\bB_1,\dots,\bB_m)$ and output symbols $\bY$. The mutual information (MI) between $\bX$ and $\bY$ is
\begin{align}\label{eq:MI}
    I(\bY;\bX)&=I(\bY;\bB_1,\dots,\bB_m) \notag\\ &=I(\bY;\bB_1)+I(\bY;\bB_2|\bB_1)+\dots \notag\\
    &+I(\bY;\bB_m|\bB_1,\dots,\bB_{m-1}) 
\end{align}
according to the chain rule. If the conditions in all conditional MIs of \eqref{eq:MI} are neglected, the BICM capacity \cite{agrell11}
\begin{align}
     I_{\text{BMI}}&=\sum_{i=1}^{m}I(\bY;\bB_i)\leq I(\bY;\bX)\label{eq:GMI}
\end{align}
is obtained. The channel is regarded as $m$ independent bit subchannels, which can be encoded and decoded independently. BICM utilizes this concept and contains only one binary component code to protect all bit subchannels. An interleaver is added between the encoder and symbol mapper to distribute the coded bits evenly to all bit subchannels.

The Gray mapping in Section~\ref{sec:pseudo} maps each bit level independently, and is suitable to be combined with BICM. Fig.~\ref{Fig:BICM} illustrates the BICM scheme for VCs. 
The total rate of BICM for VCs is $\beta \Rc$ [bits/2D-symbol], where $\Rc$ is the code rate of the inner code. Decoding is based on bit LLRs, which is described below.

For a constellation $\Gamma$ transmitted over the AWGN channel, the max-$\log$ approximation \cite{viterbi98} of the $k$th bit after receiving a $\by^j\in\R^n$ for $j=1,\ldots,N/m$ is defined as
\begin{align}
    \text{LLR}(b_{k}|\by^j)=  -\frac{1}{\sigma^2}\left (\min_{\bx\in\Gamma^{(k,0)}}\|\by^j-\bx\|^2- \min_{\bx\in\Gamma^{(k,1)}}\|\by^j-\bx\|^2\right),\label{eq:IndependentLLR}
\end{align}
where $\sigma^2$ is the noise power per two dimensions, $\Gamma^{(k,0)}$ and $\Gamma^{(k,1)}$ are the sets of constellation points 
with 0 and 1 at bit position $k$, respectively, and $\Gamma^{(k,0)}\cup \Gamma^{(k,1)}=\Gamma$. Computing \eqref{eq:IndependentLLR} needs a full search in $\Gamma$, which is infeasible for very large constellations. In \cite{ourTC}, an LLR approximation method based on importance sampling is proposed and exemplified for very large VCs based on the lattice partition $\Z^n/\Lambdas$ for the AWGN channel. The idea is to only search from a small portion of the whole constellation, which is called ``importance set''. In this paper, instead of searching from a subset of the VC, we further reduce the complexity of the approximation in \cite[Eq. (33)]{ourTC} by searching from a finite number of lattice points from $\Z^n-\ba$ that are inside a ``Euclidean ball'' centered at $\lfloor\by^j+\ba\rceil$, i.e.,
\begin{align}\label{eq:ball}
    \B(\by^j,R^2)\triangleq\{\be:\|\be+\ba-\left \lfloor \by^j+\ba  \right \rceil\|^2 \leq R^2, \be+\ba \in \Z^n\},
\end{align}
where $\lfloor\cdot\rceil$ represents rounding a vector to its nearest integer vector and $R^2\ge 0$ is the squared radius of the Euclidean ball. When the SNR is low, the points in the Euclidean ball might partially or fully fall outside of the constellation boundary. Compared with the method in \cite[Eq. (33)]{ourTC}, where the importance set is defined as $\B(\by^j,R^2)\cap \Gamma$, searching among all points from $\B(\by^j,R^2)$ might return a point outside $\Gamma$, which causes a loss in decoding performance. However, the computation complexity is reduced a lot since no closest lattice point quantizer needs to be applied to all points in $\B(\by^j,R^2)$ in order to determine the intersection with $\Gamma$.

The bit labels of the points in the Euclidean ball ${\be\in\B(\by^j,R^2)}$ can be obtained by 
\begin{align}
    f^{-1}_{\text{BRGC}}\left(w(\be),\bh)\right).
\end{align}
Then for each bit position $k$, $\B(\by^j,R^2)$ can be divided into two subsets $\B(\by^j,R^2)=\B^{(k,0)}(\by^j,R^2)\cup\B^{(k,1)}(\by^j,R^2)$, containing points within $\B(\by^j,R^2)$ with 0 and 1 at bit position $k$, respectively.
Thus, the approximated max-$\log$ LLRs $\bl^j$ of the $j$th channel realization $\by^j$ contain $m$ independent values $l_k^j$ for $k=1,\ldots,m$. The LLR of the $k$th bit is computed as
\begin{align}\label{eq:LLR_EUball}
    l_k^j=&-\frac{1}{\sigma^2}\biggl(\min_{\be\in\B^{(k,0)}(\by^j,R^2)}\|\by+\ba-\be\|^2 \notag \\&- \min_{\be\in\B^{(k,1)}(\by^j,R^2)}\|\by+\ba-\be\|^2\biggr).
\end{align}
If either $\B^{(k,0)}(\by^j,R^2)$ or $\B^{(k,1)}(\by^j,R^2)$ is empty, then the corresponding minimum in \eqref{eq:LLR_EUball} is set to a large default value $r>R^2$. Here only integer values of $R^2$ are considered, because the ${\| \be+\ba-\lfloor \by^j+\ba \rceil \|^2}$ in \eqref{eq:ball} is always an integer. The choice of $R^2$ involves a trade-off between computation complexity and decoding performance. For high-dimensional VCs, the LLR approximation can have high complexity when the Euclidean ball contains a larger number of points. Given $R^2$, $r$ can be roughly optimized by testing which value gives the best decoding performance.

\subsection{MLCM for VCs with SP mapping}
Denoting the terms of \eqref{eq:MI} as $I_1,\ldots,I_m$, a channel can be regarded as $m$ virtual independent ``equivalent subchannels'' with MIs
\begin{align}\label{eq:Ii}
    I_k=I(\bY;\bB_k|\bB_1,\dots,\bB_{k-1})
\end{align}
for $k=1,\ldots,m$. This concept directly implies an MLCM scheme proposed by Imai \emph{et al.} in \cite{imai77}, where the bit subchannels are protected unequally with different component channel codes and a multistage decoder decodes the bits successively from $\bB_1$ to $\bB_m$ provided that the previous bits are given. Practical design rules of the code rates can be found in \cite{wachsmann99}. The suitable labeling for MLCM is Ungerboeck's SP labeling. 

Fig.~\ref{Fig:MLCM} shows an MLCM scheme for VCs, which contains $q$ component codes with code rates $\Rc^i$ for $i=1,\ldots,q$ and the same codeword length $N$ to protect the first $np$ bit levels of the VC symbols, and the last $m-np$ bit levels remain uncoded. Thus, the MLCM for VCs has a total rate of
\begin{align}
    R_{\text{tot}}= \frac{\sum_{i=1}^qk_i\Rc^i+(m-np)}{n/2} \text{[bits/2D-symbol].}
\end{align}
The transmitter forms $m$ bits $\bb^{j}=(\bb_1^j,\ldots,\bb_q^j,b_{np+1}^j,\dots,b_m^j)$ for $j=1,\ldots,N$, where $\bb_i^j=(b_i^{k_i(j-1)+1},\ldots,b_i^{k_ij})$ are the $(k_i(j-1)+1)$th to $k_ij$th bits of $\bb_i$ for $i=1,\ldots,q$. The $\bb_i$ is illustrated in Fig.~\ref{Fig:MLCM} with length $k_iN$ for $i=1,\ldots,q$. For the SP mapping, the VC mapper first maps $\bb^j$ to an integer by $\bu^j=f_{\text{SP}}(\bb^j)$ and then encodes the integer into a VC point by $\bx^j=g(\bu^j)$. 


At the receiver side, after getting $(\hat{\bb}_1^j,\dots,\hat{\bb}_q^j)$, the coset representative $\hat{\bc}_i^j$ is obtained according to $\bC_i$ for $i=1,\ldots,q$, and the $\hat{\bc}^j\in[\Z^n/2^p\Z^n]$ is found by $\hat{\bc}^j=\sum_{i=1}^q \hat{\bc}_i^j$. Then the estimation of the transmitted point should be found by searching a point within the subset $2^p\Z^n+\hat{\bc}^j$ that is closest to $\by^j+\ba$. This is equivalent to
\begin{align}
    \hat{\bx}^j &=\Q_{2^p\Z^n+\hat{\bc}^j}(\by^j+\ba)\notag\\
    &=2^p\left\lfloor\frac{\by^j+\ba-\hat{\bc}^j}{2^p} \right\rceil+\hat{\bc}^j-\ba.\label{eq:xhat}
\end{align}
Finally, the bit labels of $\hat{\bx}^j$ are obtained by 
\begin{align}\label{eq:bhatSP}
   \hat{\bb}^j=(\hat{\bb}_1^j,\ldots,\hat{\bb}_q^j,\hat{b}_{np+1}^j,\ldots,\hat{b}_m^j)= f_{\text{SP}}^{-1}(w(\hat{\bx}^j)),
\end{align}
where the last $m-np$ bits $(\hat{b}_{np+1}^j,\ldots,\hat{b}_m^j)$ are the estimation of the uncoded bits mapped to $\hat{\bx}^j$.

The max-log LLRs of $\bb_i^j$ contains $k_i$ LLR values independent of each other, denoted by $\bl_i^j=(l_{i,1}^j,\dots,l_{i,k_i}^j)$, which is computed by the following procedure. Given $\hat{\bb}_1^j,\dots,\hat{\bb}_{i-1}^j$, the coset representatives $\hat{\bc}_1^j,\dots,\hat{\bc}_{i-1}^j$ are directly obtained according to look-up tables $\bC_{1},\dots,\bC_{i-1}$. Then we know that the corresponding integer belongs to the lattice $\Lambda^{i-1}+\sum_{t=1}^{i-1}\hat{\bc}_t^j$. In the $i$th partition step, the coset representatives $[\Lambda^{i-1}/\Lambda^i]$ have been labeled by the look-up table $\bC_i$. Then we can divide $\bC_i$ into two subsets $\bC_i=\bC_i^{(e,0)}\cup\bC_i^{(e,1)}$, representing coset representatives having a bit 0 and 1 at the $e$th bit of $\bb_i$, respectively. For all $\hat{\bc}_i^j\in\bC_i^{(e,0)}$, we find the closest point to $\by+\ba$ from the lattice $\Lambda^i+\sum_{t=1}^{i-1}\hat{\bc}_t^j+\hat{\bc}_i^j$, and denote all such closest points as the set
\begin{align}
    \mathcal{Z}_i^{(e,0)}=\{&\bz=\Q_{\Lambda^i+\sum_{t=1}^{i}\hat{\bc}_t^j}(\by+\ba):\hat{\bc}_i^j \in \bC_i^{(e,0)}\}.
\end{align}
The set $\mathcal{Z}_i^{(k,1)}$ is defined analogously. Then the max-log LLR of the $e$th bit of $\bb_i^j$ can be approximated as 
\begin{align}
    &\text{LLR}\left(b_i^j|\by, \hat{\bb}_1^j,\dots,\hat{\bb}_{i-1}^j\right)\approx l_{i,e}^j\notag\\
    &=-\frac{1}{\sigma^2}\left (\min_{\bz\in\mathcal{Z}_i^{(e,0)}}\|\by+\ba-\bz\|^2- \min_{\bz\in\mathcal{Z}_i^{(e,1)}}\|\by+\ba-\bz\|^2\right).\label{eq:LLR_MLC}
\end{align}
The computation complexity of the LLRs in MLCM depends on the partition orders $k_i$, which is much lower than the complexity of \eqref{eq:LLR_EUball} in BICM. However, MLCM uses $q$ component codes, which adds complexity and delay compared with BICM.
\begin{table}[tbp]
  \renewcommand{\arraystretch}{1.4}
  \renewcommand{\tabcolsep}{3.2pt}
  \caption{The considered VCs and TDHQ formats in the simulation.}
  \label{tab:VCs}
  \centering
  \begin{tabular}{c c c c c c}
    \hline 
    Name &$n$& $\Lambda/\Lambdas$ &  $M$  & $m$ & $\beta$ \\
    \hline \hline
    $E_8^{24}$ &8& $\Z^8/8E_8$ & $16,777,216$& 24 &6\\
    $\Lambda_{24}^{72}$ &24 &$\Z^{24}/2\Lambda_{24}\bR_{24}$ & $\approx4.7\times10^{21}$& 72 &6\\
    $E_8^{32}$ &8& $\Z^8/16E_8$ & $\approx4.3\times10^{9}$& 32 &8\\
    $\Lambda_{24}^{96}$ &24& $\Z^{24}/4\Lambda_{24}\bR_{24}$ & $\approx7.9\times10^{28}$& 96 &8\\
    $E_8^{40}$ &8& $\Z^8/32E_8$ & $\approx1.1\times10^{12}$& 40 &10\\
    $\Lambda_{24}^{120}$ &24& $\Z^{24}/8\Lambda_{24}\bR_{24}$ & $\approx1.3\times10^{36}$& 120 &10\\
    $E_8^{48}$ &8& $\Z^8/64E_8$ & $\approx2.8\times10^{14}$& 48 &12\\
    $\Lambda_{24}^{144}$ &24& 16$\Z^{24}/4\Lambda_{24}\bR_{24}$ & $\approx2.2\times10^{43}$& 144 &12\\
    $\Lambda_{16}^{76}$ &16& $\Z^{16}/16\Lambda_{16}$ & $\approx7.9\times10^{22}$ & 76 &9.5\\
    $\Lambda_{16}^{92}$ &16& $\Z^{16}/32\Lambda_{16}$ & $\approx5.0\times 10^{27}$ & 92 &11.5\\
    \hline
    Name &  $t_1,t_2$ &$M_1,M_2$  & 
    $m_{\text{QAM}}$ & $\beta_{\text{QAM}}$ \\
    \hline \hline
    TDHQ1 & 4, 4 &512,1024 &76 & 9.5\\
    TDHQ1 & 4, 4 &2048,1096 &92 & 11.5\\
    \hline
  \end{tabular}
\end{table}

\begin{table*}[tbp]
  \renewcommand{\arraystretch}{1.3}
  \renewcommand{\tabcolsep}{1.2pt}
  \caption{The parameters of the considered CM schemes in simulation.}
  \label{tab:CMparameters}
  \centering
  \begin{tabular}{c c c c c c c c c}
    \hline 
    Constellation & Mapping & $\beta$ & CM & Partition chain&$\#$LDPC codes & Code rates & $\#$Coded bit levels/$m$ & $R_{\text{tot}}$ [2D-symbol] \\
    \hline \hline 
    $E_8^{24}$ & Gray & 6& BICM& -& 1 &$\Rc=8/9$ &24/24 & 5.33\\
    64-QAM & Gray & 6& BICM &-& 1 &$\Rc=8/9$ &6/6 & 5.33\\
    $E_8^{24}$ & hybrid & 6& MLCM &$\Z^{8}/2\Z^{8}/E_{8}^{24}$& 1 &$\Rc^1=2/3$ &8/16 & 5.33\\
    $\Lambda_{24}^{72}$ & hybrid & 6& MLCM &$\Z^{24}/2\Z^{24}/\Lambda_{24}^{72}$& 1 &$\Rc^1=2/3$ &24/48 & 5.33\\
    64-QAM & hybrid & 6& MLCM &-& 1 &$\Rc^1=2/3$ &2/6 & 5.33\\
    \hline
    $E_8^{32}$ & Gray & 8& BICM& -& 1 &$\Rc=9/10$ &32/32 & 7.2\\
    256-QAM & Gray & 8& BICM &-& 1 &$\Rc=9/10$ &8/8 & 7.2\\
    $E_8^{32}$ & hybrid & 8& MLCM &$\Z^8/2\Z^8/E_8^{32}$& 1 &$\Rc^1=3/5$ &8/24 & 7.2\\
    $\Lambda_{24}^{96}$ & hybrid & 8& MLCM &$\Z^{24}/2\Z^{24}/\Lambda_{24}^{96}$& 1 &$\Rc^1=3/5$ &24/72 & 7.2\\
    256-QAM & hybrid & 8& MLCM &-& 1 &$\Rc^1=3/5$ &2/8 & 7.2\\
    256-QAM & SP & 8& MLCM &-& 2 &$\Rc^1=1/3, \Rc^2=8/9$ &2/8 & 7.22\\
    \hline
    $E_8^{40}$ & Gray & 10& BICM& -& 1 &$\Rc=9/10$ &40/40 & 9\\
    1024-QAM & Gray & 10& BICM &-& 1 &$\Rc=9/10$ &10/10 & 9\\
    $E_8^{40}$ & hybrid & 10& MLCM &$\Z^8/2\Z^8/E_8^{40}$& 1 &$\Rc^1=1/2$ &8/40 & 9\\
    $\Lambda_{24}^{120}$ & hybrid & 10& MLCM &$\Z^{24}/2\Z^{24}/\Lambda_{24}^{120}$& 1 &$\Rc^1=1/2$ &24/120 & 9\\
    1024-QAM & hybrid & 10& MLCM &-& 1 &$\Rc^1=1/2$ &2/10 & 9\\
    \hline
    $E_8^{48}$ & Gray & 12& BICM& -& 1 &$\Rc=9/10$ &48/48 & 10.8\\
    4096-QAM & Gray & 12& BICM &-& 1 &$\Rc=9/10$ &12/12 & 10.8\\
    $E_8^{48}$ & hybrid & 12& MLCM &$\Z^8/2\Z^8/E_8^{48}$& 1 &$\Rc^1=2/5$ &8/48 & 10.8\\
    $\Lambda_{24}^{144}$ & hybrid & 12& MLCM &$\Z^{24}/2\Z^{24}/\Lambda_{24}^{144}$& 1 &$\Rc^1=2/5$ &24/144 & 10.8\\
    4096-QAM & hybrid & 12& MLCM &-& 1 &$\Rc^1=2/5$ &2/12 & 10.8\\
    $E_8^{48}$ & SP & 12& MLCM &$\Z^8/D_8/E_8\bR_8/2\Z^8/E_8^{48}$& 1 &$\Rc^1=0,\Rc^2=0,\Rc^3=4/5$ &8/48 & 10.8\\
    \hline 
    $\Lambda_{16}^{92}$ & Gray & 11.5& BICM & -&1 &$\Rc=9/10$ &92/92 & 10.35\\
    TDHQ2 & Gray & 11.5& BICM & -&1 &$\Rc=9/10$ &92/92 & 10.35\\
    $\Lambda_{16}^{92}$ & hybrid & 11.5& MLCM &$\Z^{16}/2\Z^{16}/\Lambda_{16}^{92}$& 1 &$\Rc^1=2/5$ &16/92 & 10.3\\
    TDHQ2 & hybrid & 11.5& MLCM &-& 1 &$\Rc^1=2/5$ &16/92 & 10.3\\
    
    \hline
  \end{tabular}
\end{table*}

\subsection{MLCM for VCs with hybrid mapping}
The MLCM scheme for VCs with the hybrid mapping in Section~\ref{sec:hybrid} is a special case of Fig.~\ref{Fig:MLCM} with $p=p_q$ and $k_i=n(p_i-p_{i-1})$. At time step $j=1,\ldots,N$, the VC mapper maps $m$ bits $\bb^j=(\bb_1^j,\ldots,\bb_q^j,b_{np+1}^j,\ldots,b_m^j)$ to an integer $\bu^j =f_{\text{H}}(\bb^j)$ and then maps $\bu^j$ to a VC point $\bx^j=g(\bu^j)$. At the receiver side, successive decoding is performed based on $\by^j$ and all previous bits $(\hat{\bb}_1^j,\dots,\hat{\bb}_{i-1}^j)$ for decoder $i$. After decoding $(\hat{\bb}_1^j,\dots,\hat{\bb}_{q}^j)$, the coset representative $\hat{\bc}_i^j$ is obtained by \eqref{eq:b2c} for $i=1,\ldots,q$. The estimation of the coset representative of the partition $\Z^n/2^p\Z^n$ is calculated as $\hat{\bs}^j=\sum_{i=1}^q \hat{\bc}_i^j$. The estimation of the transmitted VC point $\hat{\bx}^j$ is decoded by \eqref{eq:xhat}, where $\hat{\bc}^j$ is replaced by $\hat{\bs}^j$. Finally, the estimation of bit labels of $\hat{\bx}^j$ is obtained by 
\begin{align}\label{eq:bhatH}
   \hat{\bb}^j=(\hat{\bb}_1^j,\ldots,\hat{\bb}_q^j,\hat{b}_{np+1}^j,\ldots,\hat{b}_m^j)= f_{\text{H}}^{-1}(w(\hat{\bx}^j)),
\end{align}
where the last $m-np$ bits $(\hat{b}_{np+1}^j,\ldots,\hat{b}_m^j)$ are the estimation of the uncoded bits mapped to $\bx^j$. The hybrid CM scheme for VCs has a total rate of
\begin{align}
    R_{\text{tot}}= \frac{\sum_{i=1}^q n(p_i-p_{i-1})\Rc^i+(m-np)}{n/2} \text{[bits/2D-symbol].}
\end{align}

The max-log LLRs of $\bb_i^j=(b_{i,1}^j,\dots,b_{i,n(p_i-p_{i-1})}^j)$ contain $n(p_i-p_{i-1})$ independent LLR values for $i=1,\ldots,q$, denoted by $\bl_i^j=(l_{i,1}^j,\dots,l_{i,n(p_i-p_{i-1})}^j)$ and calculated as follows. Given the previous estimated bits $\hat{\bb}_1,\dots,\hat{\bb}_{i-1}$, the coset representatives $\bc_t$ for $t=1,\ldots,i-1$ are obtained by \eqref{eq:b2c}. If $|2^{p_{i-1}}\Z^n/2^{p_i}\Z^n|$ is not a very large number, the max-log LLR of the $e$th bit of $\bb_i^j$, denoted by $l_{i,e}^j$, can be calculated using \eqref{eq:LLR_MLC} with ${\Lambda^{i-1}=2^{p_{i-1}}\Z^n}$ and ${\Lambda^i=2^{p_i}\Z^n}$. If $|2^{p_{i-1}}\Z^n/2^{p_i}\Z^n|$ is large, then $l_{i,e}^j$ can be calculated by enumerating a scaled Euclidean ball centered at the closest lattice point of $2^{p_{i-1}}\Z^n$ to $\by$, i.e.,
\begin{align}
    \D(\by,R^2)\triangleq\biggl\{&\be:\|\be+\ba-\left \lfloor \frac{\by+\ba}{2^{p_{i-1}}}  \right \rceil \cdot 2^{p_{i-1}} \|^2 \leq 2^{2p_{i-1}}R^2, \notag \\ &\be+\ba \in 2^{p_{i-1}}\Z^n\biggr\},\label{eq:D}
\end{align}
which consists of two subsets $\D(\by,R^2)=\D(\by,R^2)^{(e,0)}\cup\D(\by,R^2)^{(e,1)}$, representing points with $0$ and $1$ at the $e$th bit of $\bb_i$, respectively. Then $l_{i,e}^j$ is computed as
\begin{align}\label{eq:LLR_Hybrid}
        l_{i,e}^j=&-\frac{1}{\sigma^2}\biggl(\min_{\be\in\D^{(e,0)}(\by^j,R^2)}\|\by+\ba-\be\|^2 \notag \\&- \min_{\be\in\D^{(e,1)}(\by^j,R^2)}\|\by+\ba-\be\|^2\biggr).
\end{align}

It is worth noting that, when $p_i=i$ for $i=1,\ldots,q$ (i.e., the partition chain $\Z^n/2\Z^n/\dots/2^q\Z^n/\Lambdas$ is considered), setting $R^2=1$ in \eqref{eq:LLR_Hybrid} is sufficient, thanks to \eqref{eq:coset_reps} and \eqref{eq:c2b} in the hybrid labeling. The approximation complexity will be very low since $\D(\by,1)$ contains only $2n+1$ points. Also, $\D^{(e,0)}(\by,1)$ or $\D^{(e,1)}(\by,1)$ can never be an empty set for $i=1,\ldots,n$, due to \eqref{eq:coset_reps} and \eqref{eq:c2b} again.

\section{Performance analysis}

\begin{figure*}[tbp]
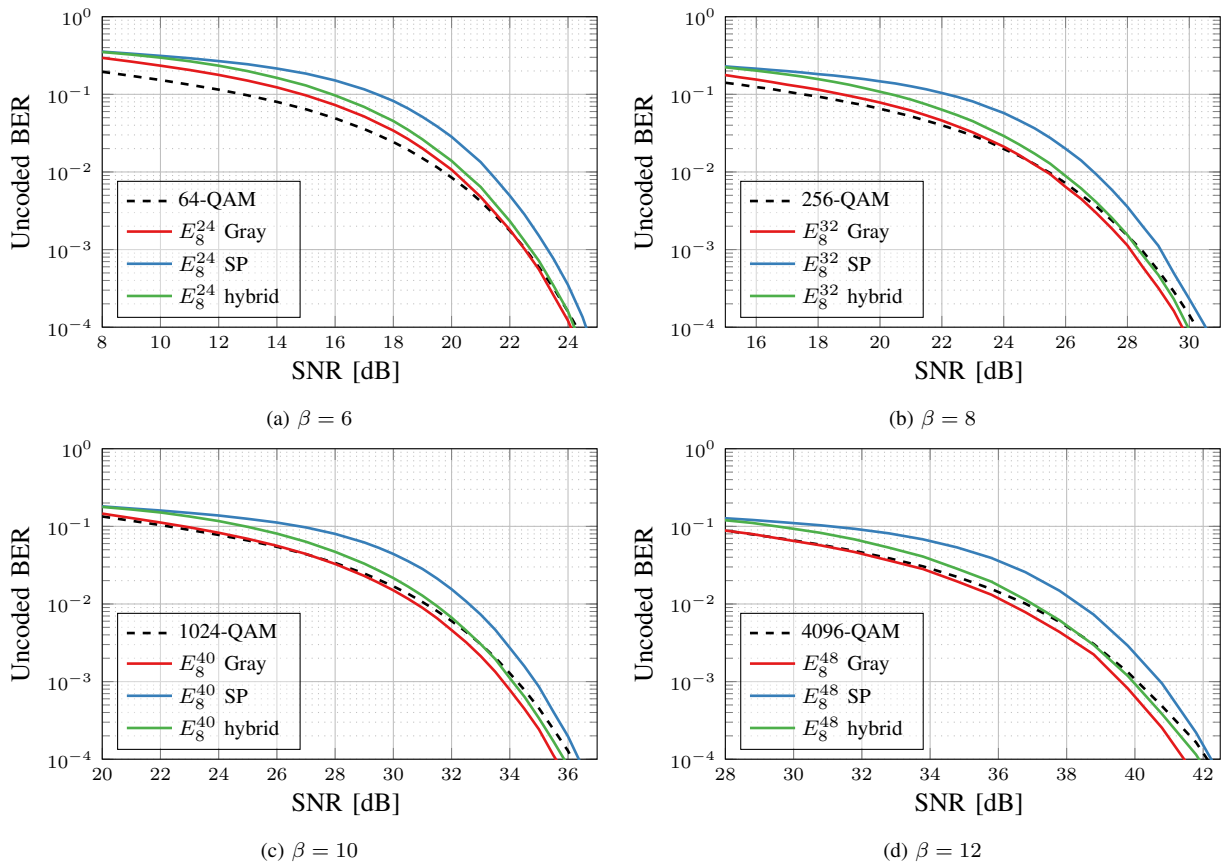

    \centering
    \begin{subfigure}{.45\linewidth}
        \input{uncoded_SE6}
        \caption{$\beta=6$}
        \label{fig:uncodedBER8D}
    \end{subfigure}
    \begin{subfigure}{.45\linewidth}
        \input{uncoded_SE8}
        \caption{$\beta=8$}
        \label{fig:uncodedBER16D}
    \end{subfigure}
    \begin{subfigure}{.45\linewidth}
        \input{uncoded_SE10}
        \caption{$\beta=10$}
        \label{fig:uncodedBER16D}
    \end{subfigure}
        \begin{subfigure}{.45\linewidth}
        \input{uncoded_SE12}
        \caption{$\beta=12$}
        \label{fig:uncodedBER16D}
    \end{subfigure}
    \caption{Uncoded BER performance of 8D VCs compared with QAM at the same spectral efficiency.}
    \label{fig:uncodedBER}
\end{figure*}
\vspace{-5pt}
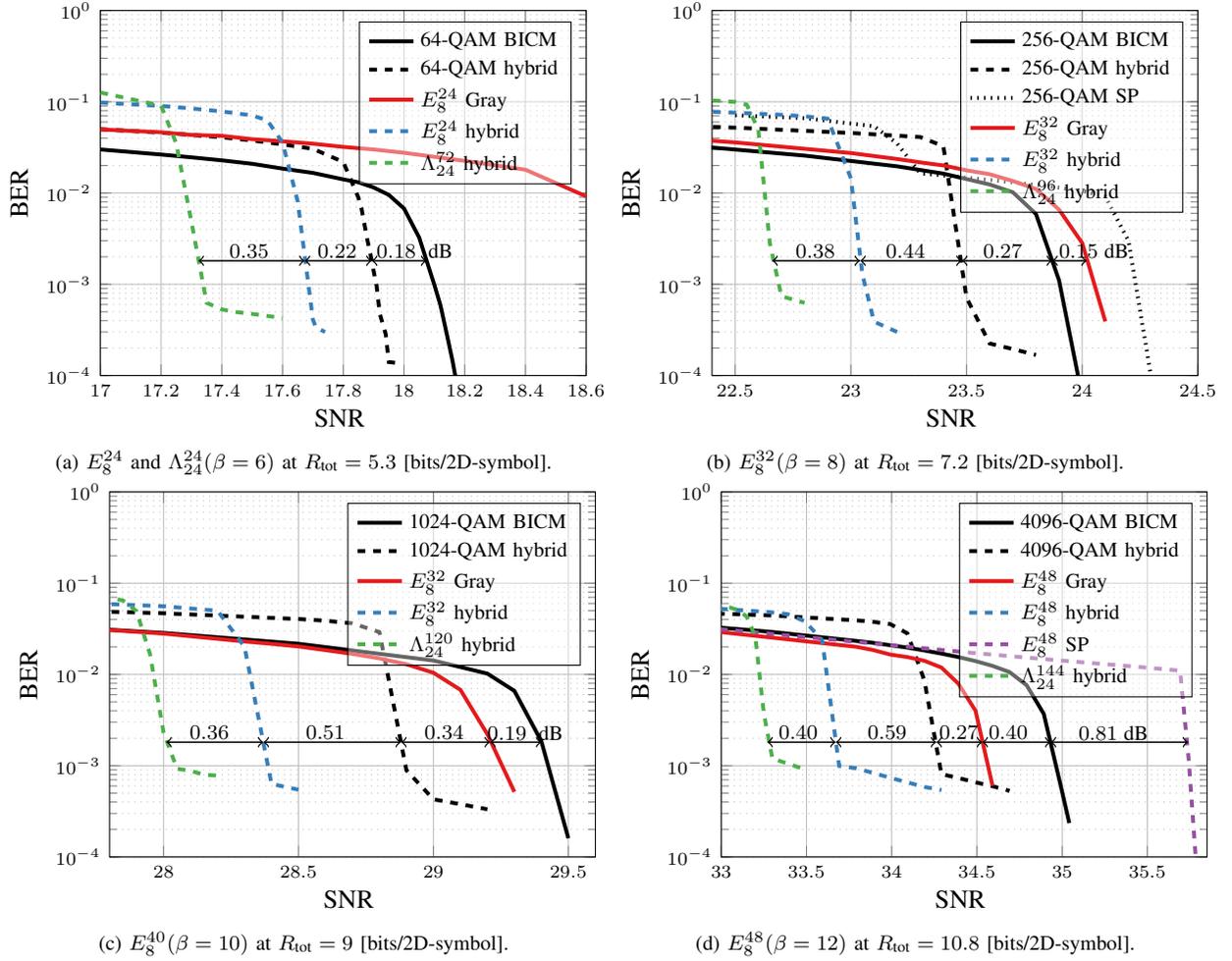
\begin{figure*}[tbp]
    \centering
    \begin{subfigure}{0.45\linewidth}
         \centering
             \begin{tikzpicture}
    \begin{semilogyaxis}[
    		xmin=17,
    		xmax=18.6,
    		ymin=1e-4, ymax=1e0,
    		xlabel={SNR},
    		ylabel={BER},
    		ylabel style={at={(axis description cs:0.03,0.5)}, anchor=north},
    		cycle list name=myCycleList,
    	    legend pos=north east,
    		legend cell align=left,
    		legend style={fill=white, fill opacity=0.4, draw opacity=1,text opacity=1},
    		ylabel style={yshift=.5cm},
    		xlabel style={xshift=-.05cm},
    		height =0.8\textwidth,
    		width=\textwidth,
    	]

        \addplot+[black,line width=1.5pt,mark=none] table[
    		x=snr,
    		y=qam,
    	] {./figures/LDPC/8D/QAM_BICM_SE6.txt};\addlegendentry{64-QAM BICM}
        \addplot+[dashed,black,line width=1.5pt,mark=none] table[
    		x=snr,
    		y=qam,
    	] {./figures/LDPC/8D/QAM_hybrid_SE6.txt};\addlegendentry{64-QAM hybrid}
        \addplot+[lines-1,line width=1.5pt,mark=none] table[
    		x=snr,
    		y=bicm,
    	] {./figures/LDPC/8D/VC_BICM_SE6.txt};\addlegendentry{$E_{8}^{24}$ Gray}
        \addplot+[dashed,lines-2,line width=1.5pt,mark=none] table[
    		x=snr,
    		y=group,
    	] {./figures/LDPC/8D/VC_hybrid_SE6.txt};\addlegendentry{$E_{8}^{24}$ hybrid}
        \addplot+[dashed,lines-3,line width=1.5pt,mark=none] table[
    		x=snr,
    		y=group,
    	] {./figures/LDPC/24D/VC_hybrid_SE6.txt};\addlegendentry{$\Lambda_{24}^{72}$ hybrid}

      \draw [<->, color=black,line width=0.2mm] (17.323,1.81e-3) to (17.673,1.81e-3);
      \node[shape=rectangle] (a) at (17.5, 2.3e-3) {\footnotesize{\textcolor{black}{$0.35\;$}}};
      
      \draw [<->, color=black,line width=0.2mm] (17.673,1.81e-3) to (17.894,1.81e-3);
      \node[shape=rectangle] (a) at (17.79, 2.3e-3) {\footnotesize{\textcolor{black}{$0.22\;$}}};
      
      \draw [<->, color=black,line width=0.2mm] (17.894,1.81e-3) to (18.074,1.81e-3);
      \node[shape=rectangle] (a) at (17.985, 2.3e-3) {\footnotesize{\textcolor{black}{$0.18\;$}}};
      \node[shape=rectangle] (a) at (18.12, 2.3e-3) {\footnotesize{\textcolor{black}{dB}}};
     
    \end{semilogyaxis}
    \end{tikzpicture}
        \caption{$E_8^{24}$ and $\Lambda_{24}^{24} (\beta=6)$ at $R_{\text{tot}}= 5.3$ [bits/2D-symbol].}
        \label{fig:codedSE6}
    \end{subfigure}
    \begin{subfigure}{0.45\linewidth}
        \centering
            \begin{tikzpicture}
    \begin{semilogyaxis}[
    		xmin=22.4,
    		xmax=24.5,
    		ymin=1e-4, ymax=1e0,
    		xlabel={SNR},
    		ylabel={BER},
    		ylabel style={at={(axis description cs:0.03,0.5)}, anchor=north},
    		cycle list name=myCycleList,
    	    legend pos=north east,
    		legend cell align=left,
    		legend style={fill=white, fill opacity=0.4, draw opacity=1,text opacity=1},
    		ylabel style={yshift=.5cm},
    		xlabel style={xshift=-.05cm},
    		height =0.8\textwidth,
    		width=\textwidth,
    	]

        \addplot+[black,line width=1.5pt,mark=none] table[
    		x=snr,
    		y=qam,
    	] {./figures/LDPC/8D/QAM_BICM_SE8.txt};\addlegendentry{256-QAM BICM}
        \addplot+[dashed,black,line width=1.5pt,mark=none] table[
    		x=snr,
    		y=qam,
    	] {./figures/LDPC/8D/QAM_hybrid_SE8.txt};\addlegendentry{256-QAM hybrid}
        \addplot+[dotted,black,line width=1.5pt,mark=none] table[
    		x=snr,
    		y=sp,
    	] {./figures/LDPC/8D/QAM_SP_SE8.txt};\addlegendentry{256-QAM SP}
        \addplot+[lines-1,line width=1.5pt,mark=none] table[
    		x=snr,
    		y=bicm,
    	] {./figures/LDPC/8D/VC_BICM_SE8.txt};\addlegendentry{$E_{8}^{32}$ Gray}
        \addplot+[dashed,lines-2,line width=1.5pt,mark=none] table[
    		x=snr,
    		y=group,
    	] {./figures/LDPC/8D/VC_hybrid_SE8.txt};\addlegendentry{$E_{8}^{32}$ hybrid}
         \addplot+[dashed,lines-3,line width=1.5pt,mark=none] table[
    		x=snr,
    		y=group,
    	] {./figures/LDPC/24D/VC_hybrid_SE8.txt};\addlegendentry{$\Lambda_{24}^{96}$ hybrid}
     
      \draw [<->, color=black,line width=0.2mm] (22.66,1.81e-3) to (23.04,1.81e-3);
      \node[shape=rectangle] (a) at (22.86, 2.3e-3) {\footnotesize{\textcolor{black}{$0.38\;$}}};

      \draw [<->, color=black,line width=0.2mm] (23.04,1.81e-3) to (23.48,1.81e-3);
      \node[shape=rectangle] (a) at (23.25, 2.3e-3) {\footnotesize{\textcolor{black}{$0.44\;$}}};
      
      \draw [<->, color=black,line width=0.2mm] (23.48,1.81e-3) to (23.87,1.81e-3);
      \node[shape=rectangle] (a) at (23.67, 2.3e-3) {\footnotesize{\textcolor{black}{$0.27\;$}}};
      
      \draw [<->, color=black,line width=0.2mm] (23.87,1.81e-3) to (24.02,1.81e-3);
      \node[shape=rectangle] (a) at (24.05, 2.3e-3) {\footnotesize{\textcolor{black}{$0.15\;$dB}}};

    \end{semilogyaxis}
    \end{tikzpicture}
        \caption{$E_8^{32} (\beta=8)$ at $R_{\text{tot}}= 7.2$ [bits/2D-symbol].}
        \label{fig:codedSE8}
    \end{subfigure}
    \begin{subfigure}{0.45\linewidth}
        \centering
            \begin{tikzpicture}
    \begin{semilogyaxis}[
    		xmin=27.8,
    		xmax=29.6,
    		ymin=1e-4, ymax=1e0,
    		xlabel={SNR},
    		ylabel={BER},
    		ylabel style={at={(axis description cs:0.03,0.5)}, anchor=north},
    		cycle list name=myCycleList,
    	    legend pos=north east,
    		legend cell align=left,
    		legend style={fill=white, fill opacity=0.4, draw opacity=1,text opacity=1},
    		ylabel style={yshift=.5cm},
    		xlabel style={xshift=-.05cm},
    		height =0.8\textwidth,
    		width=\textwidth,
    	]

        \addplot+[black,line width=1.5pt,mark=none] table[
    		x=snr,
    		y=qam,
    	] {./figures/LDPC/8D/QAM_BICM_SE10.txt};\addlegendentry{1024-QAM BICM}
        \addplot+[dashed,black,line width=1.5pt,mark=none] table[
    		x=snr,
    		y=qam,
    	] {./figures/LDPC/8D/QAM_hybrid_SE10.txt};\addlegendentry{1024-QAM hybrid}
        \addplot+[lines-1,line width=1.5pt,mark=none] table[
    		x=snr,
    		y=bicm,
    	] {./figures/LDPC/8D/VC_BICM_SE10.txt};\addlegendentry{$E_{8}^{32}$ Gray}
        \addplot+[dashed,lines-2,line width=1.5pt,mark=none] table[
    		x=snr,
    		y=group,
    	] {./figures/LDPC/8D/VC_hybrid_SE10.txt};\addlegendentry{$E_{8}^{32}$ hybrid}
      \addplot+[dashed,lines-3,line width=1.5pt,mark=none] table[
    		x=snr,
    		y=group,
    	] {./figures/LDPC/24D/VC_hybrid_SE10.txt};\addlegendentry{$\Lambda_{24}^{120}$ hybrid}

      \draw [<->, color=black,line width=0.2mm] (28.01,1.81e-3) to (28.37,1.81e-3);
      \node[shape=rectangle] (a) at (28.17, 2.3e-3) {\footnotesize{\textcolor{black}{$0.36$}}};
      
      \draw [<->, color=black,line width=0.2mm] (28.37,1.81e-3) to (28.88,1.81e-3);
      \node[shape=rectangle] (a) at (28.6, 2.3e-3) {\footnotesize{\textcolor{black}{$0.51$}}};
      
      \draw [<->, color=black,line width=0.2mm] (28.88,1.81e-3) to (29.21,1.81e-3);
      \node[shape=rectangle] (a) at (29.04, 2.3e-3) {\footnotesize{\textcolor{black}{$0.34$}}};
      
      \draw [<->, color=black,line width=0.2mm] (29.21,1.81e-3) to (29.4,1.81e-3);
      \node[shape=rectangle] (a) at (29.34, 2.3e-3) {\footnotesize{\textcolor{black}{$0.19\;\;$dB}}};

    \end{semilogyaxis}
    \end{tikzpicture}
        \caption{$E_8^{40} (\beta=10)$ at $R_{\text{tot}}= 9$ [bits/2D-symbol].}
        \label{fig:codedSE10}
    \end{subfigure}
    \begin{subfigure}{0.45\linewidth}
        \centering
            \begin{tikzpicture}
    \begin{semilogyaxis}[
    		xmin=33,
    		xmax=35.85,
    		ymin=1e-4, ymax=1e0,
    		xlabel={SNR},
    		ylabel={BER},
    		ylabel style={at={(axis description cs:0.03,0.5)}, anchor=north},
    		cycle list name=myCycleList,
    	    legend pos=north east,
    		legend cell align=left,
    		legend style={fill=white, fill opacity=0.4, draw opacity=1,text opacity=1},
    		ylabel style={yshift=.5cm},
    		xlabel style={xshift=-.05cm},
    		height =0.8\textwidth,
    		width=\textwidth,
    	]

    	\addplot+[black,line width=1.5pt,mark=none] table[
    		x expr=\thisrowno{0}+10.7918,
    		y=qam,
    	] {./figures/LDPC/8D/QAM_BICM_SE12.txt};\addlegendentry{4096-QAM BICM}
        \addplot+[dashed,black,line width=1.5pt,mark=none] table[
    		x expr=\thisrowno{0}+10.7918,
    		y=qam,
    	] {./figures/LDPC/8D/QAM_hybrid_SE12.txt};\addlegendentry{4096-QAM hybrid}
    	\addplot+[lines-1,line width=1.5pt,mark=none] table[
    		x expr=\thisrowno{0}-0.0082,
    		y=bicm,
    	] {./figures/LDPC/8D/VC_BICM_SE12.txt};\addlegendentry{$E_{8}^{48}$ Gray}
    \addplot+[dashed,lines-2,line width=1.5pt,mark=none] table[
    		x expr=\thisrowno{0}-0.0082,
    		y=group,
    	] {./figures/LDPC/8D/VC_hybrid_SE12.txt};\addlegendentry{$E_{8}^{48}$ hybrid}
        \addplot+[dashed,lines-4,line width=1.5pt,mark=none] table[
    		x expr=\thisrowno{0}-0.0082,
    		y=sp,
    	] {./figures/LDPC/8D/VC_SP_SE12.txt};\addlegendentry{$E_{8}^{48}$ SP}
       \addplot+[dashed,lines-3,line width=1.5pt,mark=none] table[
    		x =snr,
    		y=group,
    	] {./figures/LDPC/24D/VC_hybrid_SE12.txt};\addlegendentry{$\Lambda_{24}^{144}$ hybrid}
     
      \draw [<->, color=black,line width=0.2mm] (33.27,1.81e-3) to (22.88+10.7918,1.81e-3);
      \node[shape=rectangle] (a) at (33.47, 2.3e-3) {\footnotesize{\textcolor{black}{$0.40\;$}}};

      \draw [<->, color=black,line width=0.2mm] (22.88+10.7918,1.81e-3) to (23.47+10.7918,1.81e-3);
      \node[shape=rectangle] (a) at (23.2+10.7918, 2.3e-3) {\footnotesize{\textcolor{black}{$0.59\;$}}};
      
      \draw [<->, color=black,line width=0.2mm] (23.47+10.7918,1.81e-3) to (23.74+10.7918,1.81e-3);
      \node[shape=rectangle] (a) at (23.62+10.7918, 2.3e-3) {\footnotesize{\textcolor{black}{$0.27\;$}}};
      
      \draw [<->, color=black,line width=0.2mm] (23.74+10.7918,1.81e-3) to (24.14+10.7918,1.81e-3);
      \node[shape=rectangle] (a) at (23.91+10.7918, 2.3e-3) {\footnotesize{\textcolor{black}{$0.40\;$}}};
      
      \draw [<->, color=black,line width=0.2mm] (24.14+10.7918,1.81e-3) to (35.74,1.81e-3);
      \node[shape=rectangle] (a) at (35.3, 2.3e-3) {\footnotesize{\textcolor{black}{$0.81\;$dB}}};
     
    \end{semilogyaxis}
    \end{tikzpicture}
        \caption{$E_8^{48} (\beta=12)$ at $R_{\text{tot}}= 10.8$ [bits/2D-symbol].}
        \label{fig:codedSE12}
    \end{subfigure}
    \caption{Coded BER performance of 8D and 24D VCs compared with QAM at the same total rate. Solid lines represent BICM performance and dashed lines represent MLCM performance with hybrid mapping. The LLRs of VCs with BICM are calculated using \eqref{eq:LLR_EUball} with $R^2=6$ and $r=20$; the LLRs of VCs with MLCM and hybrid mapping are calculated using \eqref{eq:LLR_Hybrid} with $R^2=1$; the LLRs of VCs with MLCM and SP mapping are calculated using \eqref{eq:LLR_MLC}.}
    \label{fig:coded8D}
\end{figure*}


In this section, we present the coded BER performance in the AWGN channel for VCs with the three proposed CM schemes introduced in Section~\ref{sec:CM}, and compare them with the most commonly used benchmark at the same rate: Gray-labeled QAM with BICM \cite{caire98,ifabregas08book,stern20,stern21}. In order to see how much gain is actually from shaping, we also apply the proposed hybrid mapping in Section~\ref{sec:hybrid} to QAM and combine it with MLCM. This scheme is new, but other types of hybrid mapping for QAM with MLCM exist in the literature \cite{isaka98,yuan6g21}. However, optimizing the design of MLCM for QAM in concatenated CM schemes is not the focus of this paper.

For fairness of comparison, $n/2$ 2D QAM formats are multiplexed in the time domain to fill the same number of dimensions and to achieve the same uncoded spectral efficiencies as VCs. The traditional way of realizing non-integer spectral efficiencies for QAM formats is through time-domain hybrid QAM (TDHQ) \cite{cho19,zhugeofc13,zhoujlt13,zhoucm13}. To form an $n$-dimensional TDHQ format, $t_1$ and $t_2$ 2D QAM formats with cardinalities $M_1$ and $M_2$, respectively, are used, satisfying
\begin{align}
&t_1+t_2=\frac{n}{2} \notag\\&\beta_{\text{QAM}}=\beta=\frac{t_1\log_2(M_1)+t_2\log_2(M_2)}{t_1+t_2}.\notag
\end{align}
For example, one TDHQ symbol $\bx$ having the same spectral efficiency as $\Lambda_{16}^{92}$ ($\beta=11.5$ [bits/2D-symbol]) consists of $t_1=4$ $4096$-QAM and $t_2=4$ $2048$-QAM symbols:
\[\bx=
  (\underbrace{\bx_1,\bx_2,\bx_3,\bx_4}_\text{$\in$4096-QAM},\underbrace{\bx_5,\bx_6,\bx_7,\bx_8}_\text{$\in$2048-QAM}).
\]
The two constituent QAM constellations are scaled to the same minimum distance, which maximizes the minimum distance of the resulting hybrid QAM constellation for a given $n$-dimensional symbol energy $E_\text{s}$ \cite[Ch.~4.3]{principle}. 

VCs show high uncoded BER gains at high dimensions and spectral efficiencies \cite[Fig.~5]{ourjlt}. Thus, we investigate the performance of 8D, 16D, and 24D VCs with high spectral efficiencies of up to 12 bits/2D-symbol. The parameters of the considered VCs and the benchmark QAM formats are listed in Table~\ref{tab:VCs}.

Fig.~\ref{fig:uncodedBER} shows the uncoded BER for 8D VCs with three different mapping rules, compared with Gray-labeled QAM. For VCs in uncoded systems, the Gray mapping has the lowest uncoded BER among the three mappings and achieves an increasing SNR gain over QAM as $\beta$ increases, which implies that VCs with Gray mapping can outperform QAM in systems with a single HD FEC code \cite[Fig.~5]{ourjlt}. The hybrid mapping has marginal SNR gains over QAM at high SNRs, since the penalty of a non-Gray mapping for the VC almost counteracts its shaping gains. The SP mapping yields the worst performance and shows no gain over Gray-labeled QAM due to not efficient labeling. 

The performance of 8D and 24D VCs compared with QAM constellations with both BICM and MLCM in coded systems is shown in Fig.~\ref{fig:coded8D}. A set of LDPC codes from the digital video broadcasting (DVB-S2) standard \cite{dvbs2} with multiple code rates is considered as the inner code. The codeword length is $N=64800$ and $50$ decoding iterations are used. Table~\ref{tab:CMparameters} lists the parameters of the considered CM schemes in this paper. For all the VCs with hybrid mapping listed in Table~\ref{tab:CMparameters}, $p=q=1$ and $k_1=n$. If we target a BER of $1.81\times 10^{-3}$ when a zipper code \cite{sukmadji2019zipper} is used as the outer code, 8D VCs with MLCM and hybrid mapping yield an increasing SNR gain over QAM with MLCM and hybrid mapping from 0.22 to 0.59 dB as $\beta$ increases. These gains mainly come from shaping. When compared with Gray-labeled QAM with BICM, the most commonly used benchmark, $0.22+0.18=0.4$ to $0.59+0.27+0.40=1.26$ dB SNR gains are achieved by 8D VCs with MLCM and hybrid mapping. The MLCM achieves SNR gains over BICM due to its effective utilization of FEC overheads to protect the most significant bit levels. However, MLCM has a high error floor at a BER around $10^{-3}$ due to the uncoded bit levels, whereas the BICM scheme does not, as all bit levels are protected by FEC codes. 

Fig.~\ref{fig:coded8D} also shows that 8D VCs with BICM do not outperform QAM with BICM at $\beta=6$ and $\beta=8$, and start to achieve 0.19 and 0.40 dB SNR gains at $\beta=10$ and $\beta=12$, respectively. This observation is consistent with \cite[Fig. 6, Fig. 9]{ourjlt} that the GMI performance of VCs outperform QAM only at high spectral efficiencies. 

In Fig.~\ref{fig:codedSE8}, we also show the BER performance of 256-QAM with MLCM and SP mapping. In this scheme, the code rates are selected according to the ``capacity rule'' from \cite[Sec. IV-A]{wachsmann99}, i.e., the MIs of the equivalent subchannels $I_i$ defined in \eqref{eq:Ii}. As the first 2 bit levels are protected and the left 6 bit levels are uncoded, the MIs of the 8 equivalent subchannels are lower-bounded as
\begin{align}\label{eq:CMIQAM}
   &I(\bY;\bX)=I(\bY;\bB_1,\ldots,\bB_8)\notag\\
\geq&I(\bY;\bB_1)+I(\bY;\bB_2|\bB_1)+\sum_{k=3}^8I(\bY;\bB_k|\bB_1,\bB_2).
\end{align}
We define the eight conditional MI values of each bit level $I_k$ for $k=1,\ldots,8$ as the eight terms in \eqref{eq:CMIQAM} (6 terms are in the summation) and Fig.~\ref{fig:CMIQAM} shows $I_k$ for $k=1,\ldots,8$. To target a total rate of 7.2 [bits/2D-symbol], the MIs at $\text{SNR}=23.9$ dB in Fig.~\ref{fig:CMIQAM} suggests $\Rc^1=1/3$ and $\Rc^2=8/9$ approximately according to the capacity rule. The 256-QAM with MLCM and SP mapping suffers a large performance loss compared with other schemes, although two component codes are used.

In Fig.~\ref{fig:codedSE12}, we show the BER performance of $E_8^{48}$ with MLCM and SP mapping. The partition chain is $\Z^8/D_8/E_8\bR_8/2\Z^8/E_8^{48}$ with parameters $p=1, q=3, k_1=1, k_2=3$ and $k_3=4$. A bit different from \eqref{eq:MI} and \cite{wachsmann99}, since we can have multiple bits per partition level, the bits at the same partition level are considered independent of each other, and protected by the same code. Thus, the MIs of the first 8 equivalent subchannels are lower-bounded as
\begin{align} 
&I(\bY;\bB_1,\ldots,\bB_8)\geq I(\bY;\bB_1)+ I(\bY;\bB_2,\bB_3,\bB_4|\bB_1)\notag\\&+I(\bY;\bB_5,\bB_6,\bB_7,\bB_8|\bB_1,\bB_2,\bB_3,\bB_4)\\&\geq I(\bY;\bB_1)+\sum_{k=2}^4 I(\bY;\bB_k|\bB_1)+\sum_{k=5}^8 I(\bY;\bB_k|\bB_1,\dots,\bB_4)
\label{eq:conditionalMI}
\end{align}
The eight conditional MI values of each bit level $I_k$ for $k=1,\ldots,8$ are defined as the eight terms in \eqref{eq:conditionalMI}. Fig.~\ref{fig:CMIVC} shows the estimated $I_i$ for $i=1,\ldots,8$ using the method based on importance sampling proposed in \cite{ourTC,ourjlt}. Bits at the same partition level are protected by the same code. Thus, three different code rates should be used for the lattice partition $\Z^8/D_8/E_8\bR_8/2\Z^8$. For a bit level with an estimated conditional MI lower than 0.2, we do not use that subchannel to carry information, and set the code rate to 0. If we look at the MIs at $\text{SNR}=24.2$ dB in Fig.~\ref{fig:CMIVC}, the first four bits are not used to transmit information with $\Rc^1=\Rc^2=0$, and $\Rc^3=4/5$ according to the capacity rule. From Fig.~\ref{fig:codedSE12}, a 0.81 dB SNR loss is observed for $E_8^{48}$ with MLCM and SP mapping compared with 4096-QAM with BICM. This is due to the bad uncoded BER performance for VCs with SP mapping resulting from the high penalty of non-Gray labeling, and the FEC code cannot sufficiently reduce such a high uncoded BER. Thus, we do not consider SP mapping for 16D VCs in the following results.
\begin{figure}
    \centering
        \begin{tikzpicture}
    \begin{axis}[
    		xmin=22.5,
    		xmax=24.5,
    		ymin=0, ymax=1,
    		xlabel={SNR},
    		ylabel={MI [bits/2D-symbol]},
    		ylabel style={at={(axis description cs:0.03,0.5)}, anchor=north},
    		cycle list name=myCycleList,
    	    legend pos=north west,
    		legend cell align=left,
    		legend style={fill=white, fill opacity=0.4, draw opacity=1,text opacity=1},
    		ylabel style={yshift=.5cm},
    		xlabel style={xshift=-.05cm},
    		height =0.6\linewidth,
    		width=.9\linewidth,
    	]
    	\addplot+[lines-1,line width=1pt,mark=none] table[
    		x =snr,
    		y=I1,
    	] {./figures/MI/CMI-256QAM.txt};\addlegendentry{$I_1$}
    		\addplot+[lines-2,line width=1pt,mark=none] table[
    		x =snr,
    		y=I2,
    	] {./figures/MI/CMI-256QAM.txt};\addlegendentry{$I_2$}
    		\addplot+[lines-3,line width=1pt,mark=none] table[
    		x =snr,
    		y=I3,
    	] {./figures/MI/CMI-256QAM.txt};\addlegendentry{$I_3$}
    	\addplot+[lines-4,line width=1pt,mark=none] table[
    		x =snr,
    		y=I4,
    	] {./figures/MI/CMI-256QAM.txt};\addlegendentry{$I_4$}
     \addplot+[lines-5,line width=1pt,mark=none] table[
    		x =snr,
    		y=I5,
    	] {./figures/MI/CMI-256QAM.txt};\addlegendentry{$I_5$}
     \addplot+[lines-6,line width=1pt,mark=none] table[
    		x =snr,
    		y=I6,
    	] {./figures/MI/CMI-256QAM.txt};\addlegendentry{$I_6$}
     \addplot+[lines-7,line width=1pt,mark=none] table[
    		x =snr,
    		y=I7,
    	] {./figures/MI/CMI-256QAM.txt};\addlegendentry{$I_7$}
     \addplot+[lines-8,line width=1pt,mark=none] table[
    		x =snr,
    		y=I8,
    	] {./figures/MI/CMI-256QAM.txt};\addlegendentry{$I_8$}
\addplot+[dashed,black,mark=none] coordinates {(23.9,0) (23.9,1)};
     
    \end{axis}
    \end{tikzpicture}
    \caption{The MIs of all 8 bit levels for 256-QAM with the SP mapping. The $I_k$ for $k=3,\ldots,8$ overlap at approximately 1 bit/2D-symbol for all SNRs.}
    \label{fig:CMIQAM}
\end{figure}
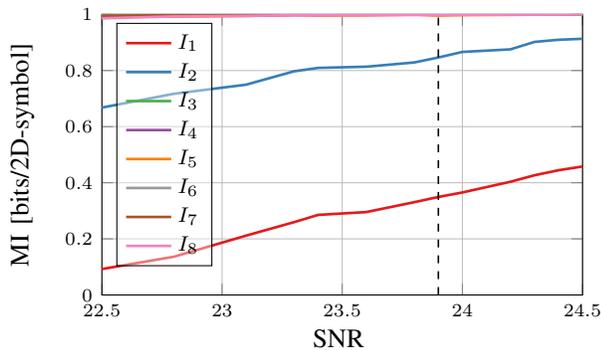

\begin{figure}
    \centering
        \begin{tikzpicture}
    \begin{axis}[
    		xmin=32.4,
    		xmax=35.5,
    		ymin=-0.1, ymax=1,
    		xlabel={SNR},
    		ylabel={MI [bits/2D-symbol]},
    		ylabel style={at={(axis description cs:0.03,0.5)}, anchor=north},
    		cycle list name=myCycleList,
    	    legend pos=north west,
    		legend cell align=left,
    		legend style={fill=white, fill opacity=0.4, draw opacity=1,text opacity=1},
    		ylabel style={yshift=.5cm},
    		xlabel style={xshift=-.05cm},
    		height =0.6\linewidth,
    		width=.9\linewidth,
    	]
    	\addplot+[lines-1,line width=1pt,mark=none] table[
    		x expr=\thisrowno{0}+10.7918,
    		y=I1,
    	] {./figures/MI/conditional.txt};\addlegendentry{$I_1$}
    		\addplot+[lines-2,line width=1pt,mark=none] table[
    		x expr=\thisrowno{0}+10.7918,
    		y=I2,
    	] {./figures/MI/conditional.txt};\addlegendentry{$I_2$}
    		\addplot+[lines-3,line width=1pt,mark=none] table[
    		x expr=\thisrowno{0}+10.7918,
    		y=I3,
    	] {./figures/MI/conditional.txt};\addlegendentry{$I_3$}
    	\addplot+[lines-4,line width=1pt,mark=none] table[
    		x expr=\thisrowno{0}+10.7918,
    		y=I4,
    	] {./figures/MI/conditional.txt};\addlegendentry{$I_4$}
     \addplot+[lines-5,line width=1pt,mark=none] table[
    		x expr=\thisrowno{0}+10.7918,
    		y=I5,
    	] {./figures/MI/conditional.txt};\addlegendentry{$I_5$}
     \addplot+[lines-6,line width=1pt,mark=none] table[
    		x expr=\thisrowno{0}+10.7918,
    		y=I6,
    	] {./figures/MI/conditional.txt};\addlegendentry{$I_6$}
     \addplot+[lines-7,line width=1pt,mark=none] table[
    		x expr=\thisrowno{0}+10.7918,
    		y=I7,
    	] {./figures/MI/conditional.txt};\addlegendentry{$I_7$}
     \addplot+[lines-8,line width=1pt,mark=none] table[
    		x expr=\thisrowno{0}+10.7918,
    		y=I8,
    	] {./figures/MI/conditional.txt};\addlegendentry{$I_8$}
\addplot+[dashed,black,mark=none] coordinates {(34.7918,0) (34.7918,1)};
     
    \end{axis}
    \end{tikzpicture}
    \caption{The MIs of the first 8 bit levels for $E_8^{48}$ with the SP mapping.}
    \label{fig:CMIVC}
\end{figure}
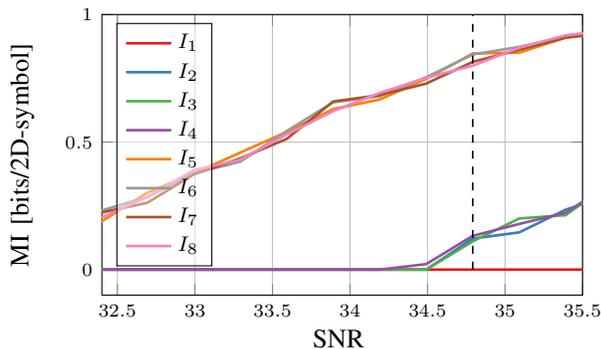
Among the 8D results, VCs with MLCM and hybrid mapping always yield the best performance. In Fig.~\ref{fig:coded8D}, we also illustrate the performance of 24D VCs with MLCM and hybrid mapping. It shows that 24D VCs can achieve 0.57 to 0.99 dB gains over QAM with MLCM and hybrid mapping at different $\beta$. When compared with QAM with BICM, up to $0.75$--$1.66$ dB gains are achieved by 24D VCs. Larger SNR gains over QAM formats are observed than in the 8D case, since 24D VCs inherently have a higher asymptotic shaping gain than 8D VCs \cite[Table I]{ourTC}.

\begin{figure*}[tbp]
    \centering
    \begin{subfigure}{0.45\linewidth}
         \centering
         \input{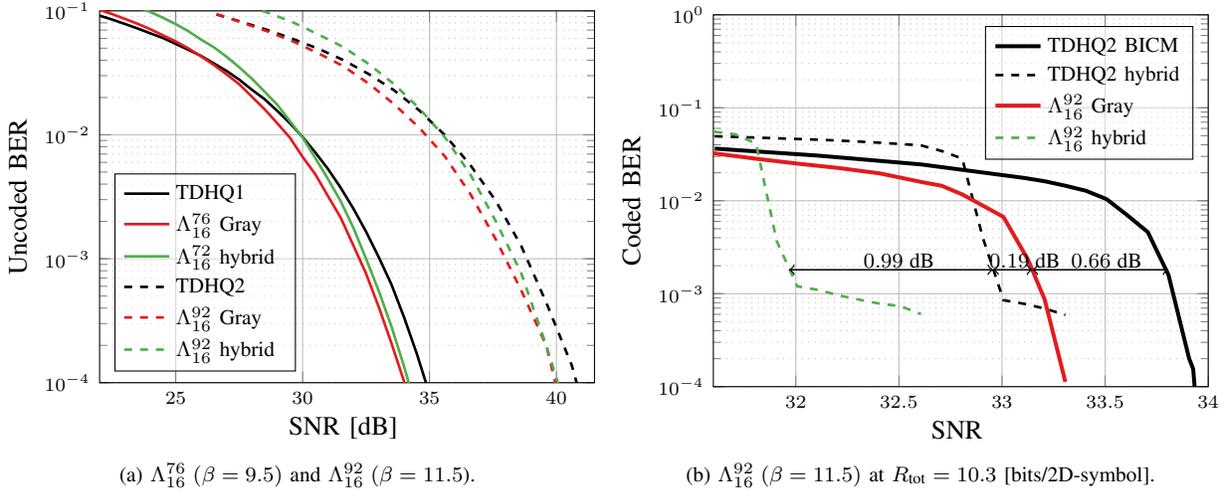}
        \caption{$\Lambda_{16}^{76}\; (\beta=9.5)$ and $\Lambda_{16}^{92}\; (\beta=11.5$).}
        \label{fig:uncoded16D}
    \end{subfigure}
    \begin{subfigure}{0.45\linewidth}
        \centering
            \begin{tikzpicture}
    \begin{semilogyaxis}[
    		xmin=31.6,
    		xmax=34,
    		ymin=1e-4, ymax=1e0,
    		xlabel={SNR},
    		ylabel={Coded BER},
    		ylabel style={at={(axis description cs:0.03,0.5)}, anchor=north},
    		cycle list name=myCycleList,
    	    legend pos=north east,
    		legend cell align=left,
    		  legend style={fill=white, fill opacity=0.4, draw opacity=1,text opacity=1},
    		ylabel style={yshift=.5cm},
    		xlabel style={xshift=-.05cm},
    		height =0.8\textwidth,
    		width=\textwidth,
    	]

          \addplot+[black,line width=1.5pt,mark=none] table[
    		x expr=\thisrowno{0}+10.607,
    		y=qam,
    	] {./figures/BER/QAM16D_BICM.txt};\addlegendentry{TDHQ2 BICM}
        \addplot+[dashed,black,line width=1pt,mark=none] table[
    		x expr=\thisrowno{0}+10.607,
    		y=qam,
    	] {./figures/BER/QAM16D_hybrid2.txt};\addlegendentry{TDHQ2 hybrid}
        \addplot+[lines-1,line width=1.5pt,mark=none] table[
    		x expr=\thisrowno{0}+10.607,
    		y=bicm,
    	] {./figures/BER/VC16D_BICM.txt};\addlegendentry{$\Lambda_{16}^{92}$ Gray}
    	\addplot+[dashed,lines-3,line width=1pt,mark=none] table[
    		x expr=\thisrowno{0}+10.607,
    		y=group,
    	] {./figures/BER/VC16D_hybrid2.txt};\addlegendentry{$\Lambda_{16}^{92}$ hybrid}

    \draw [<->, color=black,line width=0.2mm] (21.36+10.607,1.81e-3) to (22.35+10.607,1.81e-3);
      \node[shape=rectangle] (a) at (32.5, 2.3e-3) {\footnotesize{\textcolor{black}{$0.99\;$dB}}};

       \draw [<->, color=black,line width=0.2mm] (22.35+10.607,1.81e-3) to (22.54+10.607,1.81e-3);
      \node[shape=rectangle] (a) at (33.11, 2.3e-3) {\footnotesize{\textcolor{black}{$0.19\;$dB}}};
     
       \draw [<->, color=black,line width=0.2mm] (22.54+10.607,1.81e-3) to (23.2+10.607,1.81e-3);
      \node[shape=rectangle] (a) at (22.9+10.607, 2.3e-3) {\footnotesize{\textcolor{black}{$0.66\;$dB}}};

    \end{semilogyaxis}
    \end{tikzpicture}
        \caption{$\Lambda_{16}^{92}\; (\beta=11.5)$ at $R_{\text{tot}}= 10.3$ [bits/2D-symbol].}
        \label{fig:coded16D}
    \end{subfigure}
    \caption{Uncoded and coded BER performance of 16D VCs compared with TDHQ formats. Solid lines represent BICM performance and dashed lines represent MLCM performance. The LLRs of VCs with BICM are calculated using \eqref{eq:LLR_EUball} with $R^2=2$ and $r=20$; the LLRs of VCs with MLCM and hybrid mapping are calculated using \eqref{eq:LLR_Hybrid} with $R^2=1$.}
    \label{fig:16D}
\end{figure*}

For 16D VCs, which have noninteger spectral efficiencies, Fig.~\ref{fig:16D} presents the uncoded and coded BER performance compared with TDHQ formats. In Fig.~\ref{fig:coded16D}, $\Lambda_{16}^{92}$ with MLCM and hybrid mapping achieves $0.99$ dB SNR gain over TDHQ2 with MLCM and hybrid mapping. In total, it achieves up to $0.99+0.19+0.66=1.84$ dB SNR gain over TDHQ2 with BICM.

VCs with MLCM and hybrid mapping have the lowest complexity, as the scheme needs only one FEC code and only some of the bit levels are encoded. Moreover, the computation of LLRs is the fastest. In general, VCs with MLCM and SP mapping use more than one component code, which increases the complexity, and estimating the MIs of equivalent subchannels has a high complexity. VCs with BICM also use just one component code and has good performance gains at high $\beta$ thanks to the small loss in the LLR approximation. However, its LLR approximation complexity is higher than for the two MLCM schemes for VCs.

\section{Conclusion}
In this paper, we propose three CM schemes for very large MD VCs, including bit-to-integer mapping algorithms and LLR computation algorithms. This makes very large VCs adoptable in practical communication systems with SD FEC codes. Among them, one MLCM scheme for VCs with hybrid mapping has even lower complexity than BICM. The simulation results for the AWGN channel show that even with some penalty from the non-Gray labeling, VCs achieve high shaping gains over QAM with both BICM and MLCM, especially at high spectral efficiencies. Higher power gains are expected for nonlinear fiber channels, which remains as future work. 
\balance
\bibliographystyle{IEEEtran}
\bibliography{voronoi.bib}

\end{document}